\documentclass[letterpaper]{article}
\usepackage[preprint]{aaai2027}  
\usepackage[hyphens]{url}
\usepackage{graphicx}
\usepackage{natbib}
\usepackage{caption}
\usepackage{booktabs}
\usepackage{tabularx}
\newsavebox{\fitbox}
\newcommand{\fitcolumn}[1]{\sbox{\fitbox}{#1}\ifdim\wd\fitbox>\linewidth\resizebox{\linewidth}{!}{\usebox{\fitbox}}\else\usebox{\fitbox}\fi}
\newcolumntype{Y}{>{\raggedright\arraybackslash}X}
\usepackage{longtable}
\usepackage{xcolor}  
\newcommand{\answerYes}[1]{\textcolor{blue}{#1}}
\newcommand{\answerNo}[1]{\textcolor{teal}{#1}}
\newcommand{\answerNA}[1]{\textcolor{gray}{#1}}

\title{Who Benefits from Wikipedia's Growth Mentorship?\\ The Role of Newcomer Information Needs}
\author{
    Yubo Zhou,
    Pengyu Chen,
    Ceren Budak
}
\affiliations{
    School of Information, University of Michigan, Ann Arbor, Michigan, United States\\
    yubozhou@umich.edu, pyc@umich.edu, cbudak@umich.edu
}

\begin{document}

\maketitle

\begin{abstract}
Evaluations of Wikipedia's programs for new editors usually ask whether a program works on average; we ask which newcomers benefit when their question to a mentor gets a reply. English Wikipedia's Growth mentorship system assigns every new account a volunteer mentor, and about a quarter of the questions newcomers send never get a reply. Using the first questions of 34,073 newcomers (2021 to 2026), we compare newcomers who received a reply with newcomers who did not, within groups that are similar on 164 characteristics recorded before the question (propensity score stratification). (1) Does a reply help newcomers keep editing? Yes, in the short term: in the 14 days after the reply, newcomers who received one were 1.8 percentage points (pp) more likely to edit an article (22\% vs.\ an estimated 20\% without a reply) and 1.8pp more likely to edit on two or more days. Later in the first year, the only difference that remains significant after correction for multiple testing is in the share who edit an article at least once between six months and one year after the reply (+1.5pp). (2) Does the effect depend on what the newcomer asked? Yes. We sort questions by Morrison's (1993) information-seeking types, which earlier studies measured mostly through self-reports, using a language model checked against human labels. A reply has significant effects for questions about rules and norms (+3.4pp), questions that show the newcomer's earlier editing (+3.4pp) and specific how-to questions (+2.4pp). For newcomers who do not know what to work on, a reply has no detectable effect, and the effect is 4.2pp smaller than for newcomers with other questions. (3) Which replies are followed by more editing? Among newcomers who received a reply, those whose reply came sooner or stayed closer to the topic of their question edited more.
\end{abstract}

\section{Introduction}

Most newcomers to online communities leave shortly after joining. This pattern recurs across platforms, from Wikipedia's well-documented editor decline \cite{suh2009,halfaker2013} to open-source projects on GitHub where most first-time contributors never return \cite{hasan2023understanding,steinmacher2015}. Researchers and platform designers have responded with a range of retention interventions: welcome messages \cite{choi2010}, gamified tutorials \cite{narayan2017}, structured editing tasks \cite{warncke-wang2023}, dedicated Q\&A spaces \cite{morgan2013} and symbolic awards \cite{gallus2017}. Results are mixed. On Wikipedia, randomized evaluations found retention effects for the Teahouse \cite{morgan2018} and for symbolic awards to newcomers on German Wikipedia \cite{gallus2017}, while an interactive tutorial did not increase contributions \cite{narayan2017} and the Newcomer Homepage encouraged newcomers to make a first edit but did not increase retention \cite{warncke-wang2023}. Most of these evaluations ask whether the intervention works \emph{on average} and treat newcomers as one population.

Newcomers are not homogeneous in what they need, and they differ in how they edit from their first month \cite{antin2012}. A newcomer who asks ``How do I add a citation?'' seeks different information from one who asks ``What should I edit?''. \citet{morrison1993}'s framework from organizational psychology formalizes this intuition. Among the types of information that newcomers actively seek, it distinguishes \emph{technical} (how to perform tasks), \emph{referent} (what tasks to do), \emph{normative} (rules and norms) and feedback on their performance, which we call \emph{appraisal} following \citet{millerjablin1991}. Different need types predict different socialization outcomes \cite{morrison1993}, and in educational Twitter chats, whether participants return depends on their own initiative and on how the group responds to them \cite{budak2013participation}.

If newcomers need different things, the same response may help some of them and not others, and an average effect can hide this difference. Few studies of online communities estimate intervention effects separately by the type of information the newcomer asks for (Section~\ref{sec:info_seeking}).

We study which newcomers benefit from a reply using Wikipedia's Growth mentorship system. In this system, newcomers write questions directly to an assigned mentor, making each message a natural expression of the newcomer's information need. The system assigns mentors at random, so whether a newcomer's mentor tends to reply is unrelated to who the newcomer is. We compare newcomers who received a reply with similar newcomers who did not (Section~\ref{sec:identification}). The Wikimedia Growth team previously studied the effect of structured editing tasks on newcomer participation \cite{warncke-wang2023}, but excluded the mentorship module from their evaluation. We fill this gap by first estimating the overall effect of a reply on whether newcomers keep editing, and then decomposing this effect by the newcomer's information need. To do so, we classify 34,073 newcomer questions into Morrison's information-seeking categories using LLM annotation validated against two human coders (Cohen's $\kappa$ between the coders 0.84 to 0.95; between the LLM and each coder 0.68 to 0.90 on questions not used to write the prompt), and estimate category-specific effects via propensity score stratification with 164 pre-treatment covariates.

\textbf{What we measure.} Our primary outcome is whether the newcomer edits an article within 14 days of the reply. Supplementary outcomes measure how much and how widely newcomers edit and whether other editors keep their edits, and we follow newcomers for up to a year to see how long the effect lasts (Section~\ref{sec:outcomes}).

We address three research questions:
\begin{itemize}
    \item \textbf{RQ1.} Does receiving a reply increase newcomer editing? \emph{Yes, in the short term: in the 14 days after the reply, a reply raises the share of newcomers who edit an article by 1.8pp (22\% vs.\ an estimated 20\% without a reply), and the estimate holds across propensity models, numbers of strata and samples and is about half as large when both groups are measured from the same moment (Section~\ref{sec:robustness}). Later in the first year, the only difference that remains significant after correction for multiple testing is in the share who edit an article at least once in days 181--365 (+1.5pp; Section~\ref{sec:persistence}).}
    \item \textbf{RQ2.} Does the effect vary by the newcomer's information need? \emph{Yes: a reply has significant effects for questions about rules and norms (+3.4pp), questions that show the newcomer's earlier editing (+3.4pp) and specific how-to questions (+2.4pp). For newcomers who do not know what to work on, a reply has no detectable effect, and the effect is 4.2pp smaller than for newcomers with other questions (Section~\ref{sec:effects_by_type}).}
    \item \textbf{RQ3.} Which reply features are associated with higher post-reply editing? \emph{Among newcomers who received a reply, those whose reply came sooner or stayed closer to the topic of their question edited more; in a close reading of matched cases, replies followed by continued editing more often answered the question directly in the reply body (Section~\ref{sec:reply_features}).}
\end{itemize}

These findings contribute causal estimates from observational data on platform-integrated mentorship, a large-scale behavioral operationalization of Morrison's information-seeking taxonomy, and design implications for matching mentorship to newcomer needs.

\section{Related Work}

\subsection{Newcomer Retention in Wikipedia}

Wikipedia's editor population has been declining since 2007 \cite{suh2009,halfaker2013}. Multiple factors contribute to this decline: an increasingly complex rule system, aggressive quality-control mechanisms that disproportionately affect newcomers, and a cultural shift toward exclusion as the community matured \cite{halfaker2011,halfaker2013}. \citet{halfaker2011} showed that reverts by experienced editors significantly reduced newcomer retention, even when reverts were justified on quality grounds. This study also shows that the rate of such reverts increased over time. This body of work established that newcomer attrition is not simply a matter of low intrinsic motivation, but a systemic problem shaped by how the community responds to newcomers.

Several interventions have been designed to counteract this trend, with mixed results. \citet{choi2010} found that personalized welcome messages in WikiProjects slowed the decline in newcomer editing, while generic invitations had the opposite effect. \citet{morgan2018} evaluated the Teahouse (a dedicated Q\&A space for newcomers \cite{morgan2013}) and found that invited newcomers were more likely to continue editing. In a randomized field experiment on German Wikipedia, \citet{gallus2017} found that a purely symbolic award raised the share of newcomers active in the following month from 35\% to 42\%, and the difference persisted over the following year. \citet{geiger2012} randomly varied the warning messages that newcomers received after their edits were reverted and found that which message worked best depended on why the newcomer had joined. \citet{ciampaglia2015} found that MoodBar, a feature through which newcomers reported how their editing went and received replies from experienced editors, slightly increased retention. The Adopt-a-User program, a volunteer-based mentorship initiative, showed correlations between mentor--mentee communication and subsequent editing, but self-selection made causal claims difficult \cite{musicant2011}.

Other interventions did not increase retention or helped only specific groups. The Wikipedia Adventure, a gamified tutorial, did not significantly increase subsequent contributions \cite{narayan2017}. In the Wiki Education Program, where university students edit Wikipedia as class assignments, students were more likely than comparable newcomers to keep editing a year later \cite{li2020}, but the program targets one population rather than general newcomers.

Most recently, the Wikimedia Foundation's Growth team developed the Newcomer Homepage, which is a centralized landing page giving newcomers access to editing opportunities, learning materials, and suggested editing tasks \cite{warncke-wang2023}. \citet{warncke-wang2023} evaluated the Homepage and its Newcomer Tasks module through a controlled experiment across 27 Wikipedia communities ($N = 244{,}060$) and found a small increase in activation (first edit) but no effect on retention, productivity, or edit quality. Alongside these features, the Growth team also built a mentorship module into the Homepage, which assigns each newcomer a personal volunteer mentor and provides a built-in question interface that routes newcomer questions directly to their assigned mentor. This represents a shift from a shared question-and-answer space such as the Teahouse, where any experienced editor can answer, to mentorship built into the platform, where each newcomer's questions go to one assigned mentor. \citet[p.~9]{warncke-wang2023} excluded the mentorship module from their evaluation, noting that mentorship ``warrants research by itself''. This is the gap our study addresses.

Our setting also differs from that of \citet{warncke-wang2023}, who measured whether newcomers make a first edit after being given access to a tool. The newcomers we study have already contacted their mentor, and most of them (80.9\%) asked about editing Wikipedia. We measure whether a reply to that question helps them keep editing.

\subsection{Newcomer Information Seeking and Socialization}\label{sec:info_seeking}

The concept of newcomer information seeking originates in organizational psychology. \citet{morrison1993} conducted a longitudinal study of 135 newly hired accountants and identified distinct types of information that newcomers actively seek during their first months, among them \emph{technical} information (how to perform job tasks), \emph{referent} information (what tasks to do, role expectations), \emph{normative} information (organizational rules, norms, and culture) and performance feedback, which we call \emph{appraisal} information following \citet{millerjablin1991}. Morrison showed that the type of information sought predicted different socialization outcomes. For instance, technical information seeking predicted task mastery, while normative information seeking predicted cultural adaptation.

This taxonomy has been widely adopted in organizational socialization research. \citet{bauer2011} synthesized two decades of work in a comprehensive review, identifying information seeking as one of the key ``newcomer adjustment levers'' alongside socialization tactics and proactive behavior. More recently, \citet{bauer2025} published a large-scale meta-analysis of socialization programs, confirming that newcomer information seeking is one of the strongest predictors of successful organizational entry.

Despite its influence, Morrison's framework has two important limitations in its existing applications. First, as documented in reviews of the socialization literature \cite{bauer2011,bauer2025}, the taxonomy has been measured mostly through self-report surveys in which newcomers rate how often they sought each type of information, which relies on their recall. One exception, \citet{ahuja2003}, coded 673 e-mail messages of one virtual research group, 86 of them from newcomers, by the five types of \citet{morrison1993b}. Second, the framework has been applied mostly to traditional organizations (corporations, accounting firms, hospitals). Apart from \citet{ahuja2003}, Morrison's taxonomy has seen little use in online settings, and we know of few studies that connect the type of information sought to causal estimates of how the community's response affects newcomer behavior.

Two related lines of work come close but differ in important ways. \citet{choi2010} studied socialization tactics in Wikipedia, but their framework was drawn from the typology of organizational socialization tactics of \citet{vanmaanen1979} (what the organization does to newcomers), not Morrison's framework (what newcomers actively seek). \citet{ciampaglia2015} studied the MoodBar tool on Wikipedia and found that community response to newcomer messages varied by the newcomer's self-reported emotional state (happy, sad, confused), but this categorizes newcomers by affect rather than information need.

We bridge this gap by applying Morrison's taxonomy to 34,073 actual newcomer questions via LLM annotation validated against human coders, classifying each question along five binary dimensions (Section~\ref{sec:annotation}). This allows us to test whether the causal effect of a reply varies by information-seeking type.

\subsection{Causal Inference in Online Community Research}

Estimating causal effects from observational platform data is challenging because treatment assignment is typically confounded by user characteristics that also predict the outcome. Propensity score (PS) methods \cite{rosenbaum1983} address this problem, but their validity depends on having a sufficiently rich covariate set. \citet{eckles2021} benchmarked high-dimensional propensity score adjustment against a large-scale Facebook experiment (215 million observations) and showed that observational estimates converge to the experimental benchmark only when the covariate set includes rich behavioral features rather than just demographics. PS methods have since been applied successfully in platform studies \cite{maldeniya2020,mendelsohn2023}. Within Wikipedia, prior causal identification has relied on randomized invitations \cite{morgan2018} or controlled experiments \cite{warncke-wang2023}. Our setting is not randomized. Mentors are assigned at random, so the mentor's reply habits are unrelated to the newcomer, and we adjust for differences between newcomers with 164 covariates measured before the question, covering editing history, question text and time.

We use stratification rather than matching because it retains all observations and provides transparent within-stratum balance checks. Following the guidance of \citet{eckles2021}, our PS model includes 164 covariates, including the newcomer's editing history and the text of the question (Section~\ref{sec:covariates}). We assess robustness by varying the number of strata, trimming extreme propensity scores, testing alternative outcome windows, and computing E-values \cite{vanderweele2017}.

\section{Data}

\subsection{Wikipedia's Growth Mentorship System}\label{sec:growth}

Wikipedia's Growth team, a Wikimedia Foundation product team focused on newcomer onboarding, developed and deployed a mentorship module on English Wikipedia. In this system, every newly registered account  is automatically assigned an experienced volunteer editor as a personal mentor. The assignment algorithm randomly selects a mentor from a structured pool, weighted by each mentor's self-declared capacity (an integer weight of 0, 1, 2, or 4; a mentor with weight~4 receives roughly twice as many mentees as one with weight~2; weight~0 opts out of automatic assignment entirely). No newcomer characteristics enter the assignment function, making the pairing effectively random conditional on mentor availability. Mentors can also manually ``claim'' specific mentees via a dedicated interface, overriding the random assignment, but only 59 of the 37,095 newcomers  who asked a question were claimed, 19 of them before asking (these 19 are excluded).

\textbf{Entry points.} Newcomers can reach their assigned mentor through two interfaces. First, the personalized landing page for new accounts (i.e., \textbf{newcomer homepage} (\texttt{Special:Homepage})) displays a ``Your mentor'' module with the mentor's username, edit count, and a welcome message, along with a text box to ask a question. Second, while editing any article, newcomers can open a floating \textbf{help panel} that offers an ``Ask your mentor'' option alongside general editing help. In both cases, submitting a question creates a new section on the mentor's user talk page. The system tags each question with its origin (\texttt{mentorship module question} for homepage submissions, \texttt{mentorship panel question} for help panel submissions) and, for help panel questions, records the article the newcomer was editing at the time. This context appears in the section header (e.g., ``Question from User on [[Article Title]]''), enabling the mentor to provide targeted advice. The mentor receives a MediaWiki notification and replies directly on their own talk page, forming an asynchronous text-based conversation.

\textbf{Rollout and mentor pool.} Every new account is assigned a mentor, but not every newcomer could see the entry points above, because the Growth team raised  the share of new accounts that saw them in phases, from 10\% in March 2022 to 100\% in February 2025. Our data span May 2021 to April 2026 (60 months) and cover all phases. Over this period 450 mentors were on the mentor list at some time, and the newcomers in our sample were assigned to 400 of them. Appendix~\ref{app:pool} gives the rollout timeline and how the mentor pool changed.

\subsection{Sample Construction}

We collected all 42,316 tagged mentorship questions that 37,095 newcomers asked on English Wikipedia from May 4, 2021 (the first question) to April 20, 2026 (the data cutoff).  Each conversation was read from the talk page as it stood at the cutoff, including sections that had since been archived to subpages, and sections deleted before the cutoff were recovered from the page history. We keep each newcomer's first question. We exclude newcomers whose account was not an ordinary new account (e.g., created by another user or automatically created on visiting from another wiki) or who were claimed by a mentor before asking, questions whose conversation could not be found or that were removed in the very next edit (mostly reverted spam), and questions that were empty or not in English, since our text features are English-only (Appendix Table~\ref{tab:sample}). The final sample comprises $N = 34{,}073$ newcomers, of whom 26,176 (76.8\%) received at least one reply and 7,897 never received one. We call the first group \emph{treated newcomers} and the second \emph{control newcomers}.  The effect estimates use the 33,518 newcomers who can be followed for 14 days before the cutoff, counted from the reply or, for control newcomers, from a comparable time defined in Section~\ref{sec:outcomes} (25,760 treated, 7,758 control), and each outcome uses only newcomers whose follow-up for that outcome is complete. Across all 42,316 questions, 26.2\% received no reply; the share was between 22\% and 28\% in every year.

\subsection{Treatment and Outcome Definitions}\label{sec:outcomes}

\textbf{Treatment.} The treatment is whether anyone other than the newcomer replied to the newcomer's first question. The question is posted on the assigned mentor's talk page, which other editors can also see, so a reply can come from someone else. The assigned mentor replied  to 95.4\% of treated newcomers, and the other 4.6\% were answered only by other mentors or editors. Comparing newcomers who heard from the assigned mentor with those who received no reply at all gives almost the same estimate as the main analysis (+1.9 against +1.8 percentage points, pp; Appendix Table~\ref{tab:mentor_reply}). We call the time from the question to the first reply the \emph{reply lag}, and its median is 5.7 hours. For control newcomers, who never received a reply, we assign a pseudo-reply time drawn from the reply lags of treated newcomers, so that both groups' 14-day windows start at a comparable delay after the question; We therefore estimate the effect of a reply on editing in the 14 days after the reply arrives; Section~\ref{sec:robustness} shows how the estimate depends on this choice.

\textbf{Outcomes.} All outcomes count only edits made after the anchor time, which is the reply for treated newcomers and the pseudo-reply time for control newcomers. An \emph{article edit} is an edit in the main namespace. We use a 14-day primary window, the window that the Newcomer Homepage experiment used for retention, chosen to focus on newcomers' early experience while capturing most newcomers who return to edit \citep{warncke-wang2023}. Longer windows, such as the three weeks to six months used for the Teahouse \citep{morgan2018}, correspond to the later periods we use in Section~\ref{sec:persistence}. The \textbf{primary outcome} is whether the newcomer makes at least one article edit within 14 days, the most direct measure of whether the reply helped the newcomer continue editing. Two supplementary outcomes also use article edits: the \emph{number of article edits} within 14 days and \emph{return within 14 days} (article edits on at least two different days within 14 days, which separates one-time edits from coming back, similar to \citealt{warncke-wang2023}). The other two supplementary outcomes count \emph{non-talk edits}, which are edits in every namespace except the talk namespaces and so also include edits to user pages, drafts and project pages. They are the \emph{number of namespaces edited} (distinct namespaces with a non-talk edit within 14 days) and \emph{active days} (days with at least one non-talk edit within 30 days). Days are calendar days in UTC. To see how long the effect of a reply lasts (Section~\ref{sec:persistence}), we also measure whether the newcomer makes at least one, and at least five, article edits in each of four later, non-overlapping periods: days 15--28, 29--60, 61--180 and 181--365 after the anchor. Appendix Table~\ref{tab:outcomes} summarizes them.

\subsection{Question Type Annotation}\label{sec:annotation}

We classify each newcomer question along five binary dimensions grounded in the newcomer information-seeking framework of \citet{morrison1993} (Table~\ref{tab:qtypes}).

\begin{table}[t]
\centering
\small
\fitcolumn{%
\begin{tabular}{llp{3.5cm}r}
\toprule
Dim. & Name & Definition & Prev. \\
\midrule
Q1 & Substantive & Contains a question/request related to Wikipedia editing & 80.9\% \\
Q2 & Referent & Does not know what to work on & 13.5\% \\
Q3 & Appraisal & Seeks feedback on existing work & 5.6\% \\
Q4 & Normative & Asks about rules, policies, norms & 17.9\% \\
Q5 & Own work & Question shows the newcomer's earlier editing & 23.0\% \\
\bottomrule
\end{tabular}}
\caption{Question type dimensions and prevalence ($N=34{,}073$).}
\label{tab:qtypes}
\end{table}

When Q1=Y and Q2--Q4 are all N, we call the question a \emph{technical} question, that is, a specific how-to question such as ``How do I add an image?''. We call questions with Q1=N \emph{non-substantive} messages (e.g., greetings and off-topic messages). We call questions with Q2=Y \emph{referent} questions, Q3=Y \emph{appraisal}, Q4=Y \emph{normative} and Q5=Y \emph{own-work} questions. Q5 is coded independently of Q1--Q4, so a question about rules can also show the newcomer's earlier editing.

\textbf{Codebook.} Two annotators developed the codebook on the first 40 questions of a stratified random sample of 241, in two rounds of 20 questions that they labeled and discussed together.

\textbf{Human annotation.} The two annotators then labeled the other 201 questions independently, with agreement of $\kappa$ = 0.84 to 0.95 across the five dimensions (Table~\ref{tab:agreement}).

\textbf{LLM annotation.} We wrote the prompt on the first 120 questions, using some of the first 40 as examples, and labeled the questions with DeepSeek-v4-Flash in non-thinking mode. On the last 121 questions, which were not used to write the prompt, the LLM's agreement with each annotator was $\kappa$ = 0.68 to 0.90, close to the agreement between the two annotators on the same questions ($\kappa$ = 0.77 to 0.91; Table~\ref{tab:agreement}). We used this model and prompt to label every question in the analysis.

\begin{table}[t]
\centering
\small
\fitcolumn{%
\begin{tabular}{lcccc}
\toprule
Dim. & A1--A2 & A1--A2 & A1--LLM & A2--LLM \\
     & (41--241, $n$=201) & (test, $n$=121) & (test, $n$=121) & (test, $n$=121) \\
\midrule
Q1 & 0.843 & 0.771 & 0.904 & 0.809 \\
Q2 & 0.947 & 0.900 & 0.815 & 0.800 \\
Q3 & 0.875 & 0.844 & 0.900 & 0.844 \\
Q4 & 0.885 & 0.885 & 0.754 & 0.677 \\
Q5 & 0.912 & 0.909 & 0.834 & 0.784 \\
\bottomrule
\end{tabular}}
\caption{Agreement (Cohen's $\kappa$). A1--A2: the two annotators on the 201 questions they labeled independently after building the codebook (questions 41--241), and on the test set (questions 121--241, not used to write the prompt). A1/A2--LLM: each annotator against the LLM on the test set. Q4 shows the largest gap between the LLM and the annotators.}
\label{tab:agreement}
\end{table}

\section{Causal Inference Setup}

\subsection{Identification Strategy}\label{sec:identification}

We use propensity score stratification to estimate the Average Treatment Effect on the Treated (ATT). The identifying assumption is \emph{conditional ignorability}, meaning that once we condition on the observed covariates $W$, whether a newcomer receives a reply is independent of their potential outcomes. In other words, among newcomers who look the same on the measured characteristics, who gets a reply is as good as chance.

\textbf{Newcomer-side confounding.} The main threat to this assumption is that newcomers who get a reply may differ from those who do not. For example, skilled or motivated newcomers may write clearer questions, which makes a reply more likely, and they would also have edited more without one. Skill and motivation are not observed directly, so we control for 164 covariates measured before the question that reflect them, including the newcomer's editing history, the question text, and the time of the question (Section~\ref{sec:covariates}).

\textbf{Mentor-side confounding.} Replies also depend on the mentor, since some mentors answer almost every question and others rarely do. Which mentor a newcomer gets, however, is decided by chance. The system gives each new account a mentor by a weighted random draw from the mentors available at the time, without using any information about the newcomer (Section~\ref{sec:growth}). As a result, the mentor's reply habits are unrelated to who the newcomer is, and they would bias the estimate only if mentors also influenced editing in ways other than replying. In the data, newcomers of the most and the least responsive mentors are alike on the measured covariates (largest standardized mean difference 0.038), so mentors do not differ in who their newcomers are. Among newcomers who got no reply, having a more responsive mentor does not go with more editing (Appendix~\ref{app:mentor_side}), so a mentor effect outside the reply, if any, is smaller than the effect of a reply. We therefore leave mentor-side variables, six measures of the assigned mentor's past activity such as the share of earlier questions the mentor answered, out of the PS model. Adding them gives a similar estimate (+1.5pp instead of +1.8pp) with a wider interval that just includes zero (Appendix Table~\ref{tab:ablation}).

\textbf{Overlap.} The comparison also needs newcomers without a reply who resemble those with one. Section~\ref{sec:estimation} shows that this holds for almost all newcomers.

\textbf{Unobserved confounders.} The covariates cannot capture everything about a newcomer, such as how much time they have to edit. We therefore report the E-value \cite{vanderweele2017}, which says how strongly an unmeasured factor would have to be related to both receiving a reply and editing to explain away the estimate (Section~\ref{sec:robustness}).

\subsection{Covariates}\label{sec:covariates}

The 164 covariates fall into six blocks, all measured before the question: the newcomer's editing history (32 variables, e.g., total edits, revert rate, account age), features of the question text (37, e.g., length, sentiment, politeness), Perspective API toxicity scores (8), the five Morrison question types, 20 principal components of the question embedding, and 62 temporal and other controls (59 year-month indicators, whether the question was asked from an article, and 2 missing-data indicators). Appendix Table~\ref{tab:covariates} summarizes the blocks and Table~\ref{tab:covariates_full} lists every variable.

\subsection{Estimation}\label{sec:estimation}

We estimate each newcomer's propensity score $P(D=1 \mid W)$, the probability of receiving a reply ($D = 1$) given the 164 covariates $W$, with L2-regularized logistic regression ($C = 1$, the inverse of the regularization strength). Each newcomer's score comes from a model fit on the other four of five folds of the data (5-fold cross-fitting). We divide the sample into $K = 10$ strata by propensity score, compute the difference between treated and control means within each stratum, and average these differences with weights equal to each stratum's share of the treated newcomers; this average is the ATT. When we report a control mean, we weight control newcomers so that their distribution across strata matches that of treated newcomers, and we call these \emph{reweighted controls}. Confidence intervals are 95\% percentile intervals from 500 bootstrap resamples of newcomers; a full bootstrap that re-estimates the propensity score and the strata in every resample gives intervals of almost the same width (Appendix Table~\ref{tab:full_boot}). Other propensity models give the same conclusion. Making the regularization up to 100 times stronger or 10 times weaker ($C$ from 0.01 to 10) leaves the AUC at 0.626 and the ATT at 1.8 to 1.9pp. Random forests and gradient boosting predict no better (AUC 0.616 to 0.620), balance the covariates less well (after stratification, their largest standardized mean difference, SMD, is 0.051 to 0.065, against 0.027 for logistic regression; see Diagnostics), and give ATTs of 1.7 to 2.3pp with every interval excluding zero (Appendix Table~\ref{tab:ps_models}).

\textbf{Diagnostics.} We assess balance and overlap using two standard checks:

\begin{enumerate}
    \item \textbf{Overlap} (Appendix Figure~\ref{fig:ps}). The PS model predicts who receives a reply only weakly (AUC = 0.626, where 0.5 is no better than chance and 1 is perfect prediction). The propensity scores of the two groups overlap substantially, with the scores of control newcomers somewhat lower. Only 2 newcomers (0.01\%) have scores outside the range covered by both groups (the common support).

    \item \textbf{Covariate balance} (Appendix Figure~\ref{fig:love}). After stratification, every covariate has an absolute standardized mean difference (SMD) below 0.1, the conventional threshold for acceptable balance. In the raw data, 9 covariates exceed this threshold (max = 0.40); after stratification the largest is 0.027. 
\end{enumerate}

\section{Results: Overall Effect of a Reply}\label{sec:overall}

\subsection{Main Results}

\begin{table}[t]
\centering
\small
\setlength{\tabcolsep}{3.5pt}
\fitcolumn{%
\begin{tabular}{lrrrl}
\toprule
Outcome & Naive & ATT & 95\% CI & Sig \\
\midrule
\textbf{1+ edit (14d)} & +.046 & \textbf{+.018} & [+.006, +.030] & * \\
Number of article edits (14d) & +1.49 & +.784 & [$-$.063, +1.53] & \\
Namespaces edited (14d) & +.078 & \textbf{+.028} & [+.006, +.049] & * \\
Article edits on 2+ days (14d) & +.035 & \textbf{+.018} & [+.010, +.028] & * \\
Active days (30d) & +.303 & \textbf{+.135} & [+.044, +.229] & * \\
\bottomrule
\end{tabular}}
\caption{ATT estimates for the primary and supplementary outcomes ($N = 33{,}518$ newcomers with a full 14-day follow-up; 32,883 with a full 30 days for active days; 95\% bootstrap CIs). Effects are differences in proportions (+.018 = 1.8 percentage points); the number of article edits, namespaces and active days are differences in counts.}
\label{tab:att}
\end{table}

Table~\ref{tab:att} presents the main ATT estimates (see also Appendix Figure~\ref{fig:forest_overall}). The naive difference in the share of newcomers who edit an article within 14 days is 4.6pp, but more than half of this gap reflects selection, because mentors reply more often  to newcomers who would have been more likely to edit anyway. After adjustment, the ATT is +1.8pp: 22.2\% of treated newcomers edited an article within 14 days, against an estimated 20.3\% had they not received a reply. Three of the four supplementary outcomes are also significant. Newcomers who received a reply were active on more days over the following month (+0.13), edited in more namespaces (+0.028), and were more likely to come back on a second day (return within 14 days, +1.8pp). They also made more article edits (+0.78 in 14 days). Whether this interval includes zero depends on how it is computed: it does when the bootstrap resamples newcomers ([$-$0.06, +1.53]) and does not when it resamples mentors ([+0.03, +1.53]) or re-estimates the propensity score in every resample ([+0.17, +1.55]; Appendix Table~\ref{tab:full_boot}). The interval is wide because the count is highly skewed: 79\% of newcomers made no article edit in the 14 days, and the top 1\% made half of all article edits. 

\textbf{Edit quality.} More editing matters only if the edits are kept. We count an article edit as rejected if another editor reverted it within 48 hours, the window used by \citet{warncke-wang2023}. Edits that newcomers reverted themselves are not counted as rejected (Appendix~\ref{app:quality}). Among newcomers without a reply who edited an article, on average about one in five of each newcomer's article edits were rejected, and we use this level as the reference. We do not compare the share of rejected edits between the two groups' editors, because a reply changes who edits. The treated group includes  newcomers who started editing because they received a reply, so the editors in the two groups are no longer comparable. Instead, we place every newcomer, including those who did not edit, into one of three groups by their article edits in the 14 days after the anchor: (1) no article edit; (2) edits rejected no more often than the reference, that is, at least 80\% kept; (3) edits rejected more often than the reference. A reply raises the share in group 2 by 1.9pp (15.2\% among reweighted controls; 95\% CI [+1.0, +2.9]), as much as it lowers the share in group 1, and does not change the share in group 3 (5.0\%; +0.0pp [$-$0.5, +0.7]). The conclusion is the same for any cut between 50\% and 90\% kept (Appendix Table~\ref{tab:quality}).

\subsection{How Long Does the Effect Last?}\label{sec:persistence}

\begin{table}[t]
\centering
\small
\setlength{\tabcolsep}{3.5pt}
\fitcolumn{%
\begin{tabular}{lrrlr}
\toprule
Days & Treated & Ctrl. & ATT [95\% CI] & $q$ \\
\midrule
\multicolumn{5}{l}{\emph{At least one article edit}} \\
1--14 & .207 & .182 & +.025 [+.011, +.039] & -- \\
15--28 & .090 & .083 & +.006 [$-$.004, +.016] & .26 \\
29--60 & .099 & .091 & +.008 [$-$.003, +.019] & .26 \\
61--180 & .117 & .107 & +.010 [$-$.001, +.022] & .23 \\
181--365 & .092 & .077 & \textbf{+.015} [+.005, +.024] & .04 \\
\midrule
\multicolumn{5}{l}{\emph{At least five article edits}} \\
1--14 & .097 & .080 & +.016 [+.006, +.026] & -- \\
15--28 & .044 & .040 & +.004 [$-$.003, +.011] & .27 \\
29--60 & .049 & .044 & +.005 [$-$.002, +.013] & .26 \\
61--180 & .064 & .057 & +.008 [$-$.001, +.016] & .23 \\
181--365 & .051 & .045 & +.005 [$-$.003, +.013] & .26 \\
\bottomrule
\end{tabular}}
\caption{Effect of a reply on article editing in non-overlapping periods after the anchor, for the 19,512 newcomers whose anchor is at least 365 days before the data cutoff (14,858 treated). Ctrl.\ = control mean reweighted to the treated newcomers' propensity score strata. $q$: Benjamini--Hochberg over the eight later-period estimates (2,000 bootstrap draws); days 1--14 are shown for comparison and are not part of the correction. Shares and effects are proportions (+.015 = 1.5 percentage points).}
\label{tab:later}
\end{table}

A 14-day effect says little about whether newcomers stay. We therefore follow the 19,512 newcomers whose anchor is at least a year before the data cutoff, so that every period is observed for the same people. We split the year after the anchor into non-overlapping periods: the rest of the first month (days 15--28), the second month (29--60), the rest of the first half-year (61--180) and the second half-year (181--365). In each period we estimate the effect on editing an article at least once, as in the primary outcome, and at least five times, the two thresholds used by \citet{morgan2018}, with Benjamini--Hochberg $q$ values over the eight later estimates (Table~\ref{tab:later}).

In this cohort, a reply raises the share who edit an article in days 1--14 by 2.5pp (95\% CI [+1.1, +3.9]), more than the 1.8pp of the full sample, and the share who edit at least five times by 1.6pp ([+0.6, +2.6]). The later differences are smaller: seven of the eight later estimates are between +0.4 and +1.0pp, and their intervals include zero. The exception is editing at least once in days 181--365 (+1.5pp, 95\% CI [+0.5, +2.4], $q = .04$); for at least five edits in the same period the estimate is +0.5pp ($q = .26$). Few newcomers edit at all after the first month: in both groups the share who edit an article falls from 21--23\% in the first month to under 10\% in the second and about 3\% in the twelfth (Appendix Figure~\ref{fig:monthly}). As a check, we also let each period use every newcomer whose follow-up covers it, which gives larger samples that differ between periods; none of the six estimates for days 15--180 then passes the correction (Appendix Table~\ref{tab:later_all}).

\textbf{What kind of editors do newcomers become?} We follow the 19,512 newcomers whose anchor is at least a year before the data cutoff (the newcomers of Table~\ref{tab:later}) and count only their edits after the anchor. For each newcomer we record whether, within the year, they made at least 10 edits and at least 500 edits (the edit counts Wikipedia uses for autoconfirmed and extended confirmed accounts; we use only the counts and do not check whether an account received these rights), at least 5 edits in some 30-day period (Wikimedia's active editor), and at least 100 edits in some 30-day period (very active editor). We estimate the effect of a reply on each of the four levels as in Table~\ref{tab:later} and apply a Benjamini--Hochberg correction over the four tests. After the correction, the effects on two levels remain significant: a reply raised the share who made 10 edits by 2.1pp (95\% CI [+0.7, +3.3], $q = .006$) and the share who reached the active level by 3.0pp ([+1.5, +4.3], $q < .001$). The effects on the very active level (+0.5pp, $q = .08$) and on 500 edits (+0.4pp, $q = .06$) are not significant. When only edits in days 15--365 are counted, the effects on 10 edits (+0.9pp) and on the active level (+0.5pp) are not significant either. We also tested whether a reply changes what newcomers mainly edit after day 14 (articles, drafts, user and talk pages, other namespaces, a mix, or nothing); the six shares do not differ between the groups ($p = .25$; Appendix Tables~\ref{tab:trajectories} and \ref{tab:editor_type}).

\subsection{Robustness}\label{sec:robustness}

The primary ATT is stable across the number of strata (+1.7 to +2.0pp for $K = 5$ to $20$), PS trimming (+1.7 to +2.1pp; Appendix Table~\ref{tab:trimming}), cumulative outcome windows from 7 to 180 days (+1.6 to +2.0pp; Appendix Table~\ref{tab:window}), 16 feature ablation specifications (all but one significant; the exception adds mentor-side features, see Section~\ref{sec:identification}), and six alternative samples and treatment definitions, including requiring a reply from the assigned mentor, dropping questions asked in the last 30 days before the cutoff, and keeping only newcomers with complete text annotations (+1.8 to +2.1pp; Appendix Tables~\ref{tab:strata_sens}--\ref{tab:robust_specs} and Figure~\ref{fig:robustness}). Defining treatment as a reply within 48 hours gives +5.8pp; this compares fast replies with late or no replies, a different contrast (Appendix Table~\ref{tab:robust_specs}). The full bootstrap gives a 95\% CI of [+1.2, +3.2] for the 1.8pp effect. The E-value is 1.41 for the point estimate and 1.20 for the lower confidence limit \cite{vanderweele2017}.

\textbf{Anchor time.} Control newcomers never receive a reply, so we have to decide when their 14-day window starts. The decision matters because \textbf{newcomers edit most in the hours right after they ask} and less as time goes on. A window for control newcomers that starts early therefore counts more editing than one that starts late. If the window for every control newcomer starts at the moment of the question, treated newcomers appear 8.1pp less likely to edit an article within 14 days than control newcomers. If it starts at the 75th percentile of treated reply lags (19.2 hours after the question), treated newcomers appear 3.7pp more likely (Appendix Table~\ref{tab:anchor}). A fair start time should count only edits made after the reply, or after the time a reply would have arrived. It should also observe the two groups at the same point after their question. Our main design does both by giving each control newcomer \textbf{a lag drawn at random from the reply lags of treated newcomers} (median 5.7 hours), so that windows in the two groups start at the same spread of delays. The random draw works only if reply lags depend neither on what newcomers do while they wait nor on who they are. In our data, newcomers who kept editing while they waited were answered about as often as newcomers with similar propensity scores who did not, the median lag is between 4.9 and 6.4 hours in every propensity stratum, and drawing each control's lag from treated newcomers in the same stratum gives +1.7pp (Appendix~\ref{app:anchor}). With this design, a reply raises the share of newcomers who edit an article within 14 days by 1.8pp.

Two other designs start both groups at the same moment and answer different questions. Starting at the question estimates the effect of eventually receiving a reply on the 14 days after asking. It counts the hours in which treated newcomers are still waiting: of the treated newcomers who edit an article in those 14 days, 57\% make their first edit before the reply arrives, when the reply cannot yet have had any effect. This design gives 0.9pp (95\% CI [$-$0.3, +2.1]). Starting 48 hours after the question compares replies within 48 hours with later replies or none, so 32\% of its controls received a reply later. It gives 0.8pp (95\% CI [$-$0.1, +1.7]), and 1.4pp (95\% CI [+0.2, +2.4]) when the controls are limited to newcomers who never received a reply (Appendix~\ref{app:anchor}). Across all landmark times we tried, from 12 to 168 hours, the estimate is +0.3 to +1.0pp, smaller than in the main design because the controls of every landmark include newcomers who were answered later (Appendix Table~\ref{tab:landmark}).

\section{Results: Effects by Question Type}\label{sec:effects_by_type}

\subsection{Motivation and Method}

The overall ATT averages across all newcomers, but newcomers with different information needs may benefit differently. We estimate a \textbf{separate PS model within each subgroup} (using the remaining 159 covariates, excluding Q1--Q5 since we condition on question type by subsetting, and $K = 5$ strata because the subgroups are smaller). This approach avoids imposing a shared PS model on subgroups with very different baseline editing rates (the share of control newcomers who edit an article within 14 days ranges from 9.3\% to 29.7\%).

\subsection{Main Results}

\begin{table}[t]
\centering
\small
\setlength{\tabcolsep}{4pt}
\fitcolumn{%
\begin{tabular}{lrrrl}
\toprule
Subgroup & $N$ & ATT & 95\% CI & Sig \\
\midrule
Full sample & 33,518 & +.023 & [+.011, +.034] & * \\
\textbf{Technical} & 15,750 & \textbf{+.024} & [+.007, +.041] & * \\
\textbf{Normative (Q4)} & 5,974 & \textbf{+.034} & [+.003, +.066] & * \\
Appraisal (Q3) & 1,864 & +.040 & [$-$.012, +.091] & \\
Referent (Q2) & 4,528 & $-$.016 & [$-$.046, +.012] & \\
Non-substantive & 6,424 & +.006 & [$-$.011, +.021] & \\
\midrule
\textbf{Own work (Q5=Y)} & 7,688 & \textbf{+.034} & [+.009, +.060] & * \\
No own work (Q5=N) & 25,830 & +.020 & [+.008, +.032] & * \\
\bottomrule
\end{tabular}}
\caption{Within-subgroup ATT by question type (primary outcome: 1+ article edit, 14d). Q1--Q4 subgroups are not mutually exclusive. ``Technical'' = Q1=Y with Q2--Q4 all N. The full-sample row uses the subgroup specification (159 covariates, $K = 5$), so it differs slightly from Table~\ref{tab:att}. Effects are differences in proportions (+.024 = 2.4 percentage points).}
\label{tab:subgroup}
\end{table}

Table~\ref{tab:subgroup} reports the within-subgroup ATT for the primary outcome (see also Appendix Figure~\ref{fig:subgroup_forest} for visualization). Appendix Table~\ref{tab:subgroup_all} gives the estimates for all six outcomes in every subgroup. To test whether the effect differs between question types, we fit one regression on all newcomers in which the effect of a reply may differ between one question type and all other questions, with propensity stratum fixed effects and a newcomer bootstrap. For referent questions, the effect of a reply is 4.2pp smaller than for all other questions (95\% CI [$-$6.9, $-$1.2]; $q = .03$ after a Benjamini--Hochberg correction over the eight comparisons). None of the other question types differs significantly from the remaining questions.

\textbf{Significant effects.} The effect on the primary outcome is significant for technical questions (+2.4pp), normative questions about rules and norms (Q4, +3.4pp), own-work questions (Q5=Y, +3.4pp) and the other questions (Q5=N, +2.0pp). Q5 is coded independently of Q1--Q4, and the difference between Q5=Y and Q5=N is not significant (+1.6pp, 95\% CI [$-$1.2, +4.4]). For return within 14 days, the four estimates are +2.3pp, +2.7pp, +2.6pp and +1.9pp, all significant except the one for normative questions.

\textbf{Nulls.} Newcomers who ask \textbf{referent questions} (Q2), who do not know what to work on, show no detectable effect ($-$1.6pp, 95\% CI [$-$4.6, +1.2]) for every number of strata ($K = 3$, 5, 7 and 10). \textbf{Non-substantive} messages (Q1=N, such as greetings and off-topic messages) also show no significant effect on the primary outcome (+0.6pp), as expected for messages such as ``Hello!'', where a reply has little reason to cause article editing. 

\textbf{Non-significant positives.} For appraisal questions (Q3), the estimate is +4.0pp, but with 1,864 newcomers the 95\% CI [$-$1.2, +9.1] includes zero.

\textbf{Sensitivity checks.} With $K$ from 3 to 10 strata (Appendix Figure~\ref{fig:strata_sensitivity}), technical,  Q5=Y and Q5=N remain significant for every $K$, normative for every $K$ except 7, the referent and non-substantive nulls remain null, and appraisal is significant only at $K = 10$.

\section{Results: What Makes a Reply Effective?}\label{sec:reply_features}

\subsection{Motivation and Design}

Having estimated the effect of receiving a reply (Sections~\ref{sec:overall}--\ref{sec:effects_by_type}), we ask: among newcomers who received a reply, which features of the reply text are associated with higher post-reply editing? This analysis is \textbf{exploratory and associational}, because reply features are post-treatment variables  and cannot be given a causal interpretation \cite{rosenbaum1984}. Its purpose is to generate hypotheses for future experimental work and to test whether different question types are sensitive to different reply features.

We describe each first reply with 19 features (Appendix Table~\ref{tab:features}). Seventeen are text features covering length, pronoun use, sentiment, structure (readability, imperatives, lists, question marks), links to Wikipedia policy and help pages, and interaction cues (greeting, gratitude, politeness). The other two are \textbf{question--reply similarity} (the cosine similarity of the question and reply embeddings, measuring how closely the reply addresses the specific question asked) and \textbf{reply lag} (hours between the newcomer's question and the first reply).

We regress the primary outcome on the standardized reply features and the same 164 pre-treatment covariates, using a linear mixed model with a random intercept for the mentor (the estimated mentor variance is close to zero, so the results equal ordinary least squares). Two further link counts, total links and wikilinks, are nearly collinear with these (variance inflation factor $\approx$ 52) and are left out. Analyses cover the treated newcomers with a full 14-day follow-up ($N = 25{,}756$) and each Morrison subgroup separately. For subgroup analyses, we run separate regressions on each subset rather than adding interaction terms to a pooled model; this allows both the text-feature coefficients and the control-variable coefficients to vary freely across subgroups, at the cost of reduced statistical efficiency from smaller within-subgroup samples.

\subsection{Full Sample Results}

\begin{table}[t]
\centering
\small
\fitcolumn{%
\begin{tabular}{lrrl}
\toprule
Feature & $\beta$ (std) & $p$ & Sig \\
\midrule
Reply lag (hours) & $-$0.033 & $<$.001 & *** \\
Question--reply similarity & +0.015 & $<$.001 & *** \\
Sentiment score & +0.008 & .009 & ** \\
Share of ``we'' words & $-$0.005 & .021 & * \\
Subjectivity & +0.006 & .014 & * \\
Has list & +0.005 & .026 & * \\
\bottomrule
\end{tabular}}
\caption{Significant reply feature associations with the primary outcome (1+ article edit, 14d). $\beta$ is the change in the probability of an article edit within 14 days for a one-standard-deviation increase in the feature. Sentiment score: VADER compound score, from $-1$ (most negative) to $+1$ (most positive). Subjectivity: TextBlob score, from 0 (objective) to 1 (subjective). Share of ``we'' words: first-person plural pronouns per word of the reply. Has list: the reply contains a bulleted or numbered list. * $p < .05$, ** $p < .01$, *** $p < .001$. Only features with $p < .05$ shown. After a Benjamini--Hochberg correction over the 19 features, only reply lag and question--reply similarity remain significant (both $q < .001$, against $q = .054$ to .081 for the other four).}
\label{tab:reply_full}
\end{table}

Faster replies are associated with higher post-reply editing, and \textbf{reply lag} is by far the strongest predictor (Table~\ref{tab:reply_full}; Appendix Figure~\ref{fig:reply_features}). Replies whose embeddings are closer to the question embedding, that is, replies that stay on the newcomer's topic, are also associated with higher post-reply editing; this \textbf{question--reply similarity} is the second strongest predictor ($\beta = +0.015$, $p < .001$). Among the remaining content features, a higher sentiment score, subjectivity and the presence of a list are modestly associated with more editing and a higher share of ``we'' words with less editing, but none of these four survives correction for multiple testing ($q = .054$ to .081).

\subsection{Subgroup Patterns}

We fit the same regression in each of the six subgroups (Appendix Table~\ref{tab:reply_subgroup_full}). Reply lag is significant in every subgroup ($\beta = -0.015$ to $-0.051$). Question--reply similarity is significant for technical and normative questions and for Q5=N, but not for referent questions, for which a reply itself has no detectable effect. A few other features are significant in single subgroups: pointing technical askers to policy pages goes with less editing ($\beta = -0.013$), more question marks go with more editing for own-work questions ($\beta = +0.013$), and a higher share of ``we'' words goes with less editing for Q5=N ($\beta = -0.005$). These subgroup coefficients are not corrected for multiple testing; with 19 features in each of six subgroups (114 tests) at the 5\% level, about six would be significant by chance even if no feature mattered.

\textbf{Beyond the first reply.} Our design uses only the first question and first reply, but many conversations continue (Appendix Table~\ref{tab:conversation}). Among treated newcomers, 17.4\% wrote back after the first reply, 11.9\% received more than one reply from a mentor, and 8.2\% later asked another question, almost always on the same mentor's page. The median conversation has two turns and is finished within a day (median 8 hours). Newcomers who wrote back were far more likely to edit an article within 14 days (46.8\% vs.\ 17.0\%; 20pp within propensity score strata). The strata match newcomers only on what was measured before the question. Whether a newcomer writes back is decided after the reply, at the same time as whether they keep editing, so the gap may reflect how engaged the newcomer was at that point rather than an effect of the conversation.

\subsection{Close Reading of Matched Cases}

The text features above describe the surface of a reply. To see whether reply content differs in ways these features miss, we read the replies in 9 groups of matched newcomers who asked similar questions and had similar editing histories but differed in whether they kept editing (54 cases; Appendix~\ref{app:close_reading}). The feature that most distinguished newcomers who kept editing was whether the reply answered the question directly in its text rather than pointing to a help page (61\% vs.\ 31\%), followed by whether it described concrete action steps (44\% vs.\ 19\%).

\section{Discussion}

\subsection{Summary of Findings}

The overall ATT of 1.8pp comes from a single asynchronous text reply. The Newcomer Homepage, evaluated in a controlled experiment, raised the share of accounts that made a first edit by a similar amount (2.5pp; \citealt{warncke-wang2023}). The 1.8pp effect corresponds to about 470 more of the 25,760 treated newcomers editing an article within 14 days. The result for return within 14 days (+1.8pp) shows that the reply does not only trigger one edit but also brings newcomers back on another day. Later in the first year, the only difference that remains significant after correction for multiple testing is in the share who edit an article at least once in days 181--365 (+1.5pp). In both groups, fewer than one in ten newcomers edit an article in any month after the first.

The effect also differs by question. It is significant for normative questions about rules and norms (+3.4pp), own-work questions (+3.4pp) and technical questions (+2.4pp). For referent questions, which come from newcomers who do not know what to work on, a reply has no detectable effect, and the effect is 4.2pp smaller than for newcomers with other questions. Among reply features, reply speed has the strongest association with later editing, followed by question--reply similarity; the remaining surface-level features (word count, sentiment, link counts) have small associations. In the close reading, replies followed by continued editing more often answered the question directly in the reply body rather than redirecting to an external resource, and detailed replies to newcomers who wanted to create pages for an organization were not followed by continued editing.

\subsection{Practical Implications}

\begin{enumerate}
    \item \textbf{Prioritize response speed.} Reply lag is associated with less post-reply editing in every subgroup and is the strongest of the 19 features in four of the six. Platform interventions that reduce response time, such as notification nudges, mentor availability dashboards, or routing unanswered questions to available mentors, are candidates for experiments that test whether faster replies increase editing.

    \item \textbf{Answer the question directly.} In the close reading (Appendix~\ref{app:close_reading}), newcomers who kept editing had more often received a reply that gave the answer in the reply itself, such as the syntax for a citation tag, rather than only a link to a help page (61\% vs.\ 31\%). With 54 cases, whether mentors should give the answer first and add links second is a hypothesis for an experiment. 

    \item \textbf{Newcomers who do not know what to work on.} A reply has no detectable effect for referent questions (Q2), and the effect is 4.2pp smaller than for newcomers with other questions. Whether these newcomers respond to other support, such as suggested articles or guided tasks, is untested.
\end{enumerate}

\subsection{Limitations}

\textbf{E-value.} The E-value for the lower confidence limit (1.20) means that an unmeasured confounder associated with both reply receipt and editing by a risk ratio of 1.20 each could move the confidence interval to include zero. The E-value depends only on the size of the estimate: a reply raises the share who edit an article from an estimated 20.3\% to 22.2\%, a risk ratio of 1.09, and an estimate of this size has an E-value of 1.41. A low E-value here reflects how small the effect of a single reply is.

\textbf{Anchor time.} Control newcomers never received a reply, so we start their 14-day window at a pseudo-reply time drawn from the reply delays of treated newcomers. This is fair only if how long a reply takes depends neither on what newcomers do while waiting nor on who they are, and we find no sign that it does (Appendix~\ref{app:anchor}). When both groups are instead measured from the same moment, the estimate is about half as large (0.8 to 0.9pp instead of 1.8pp) and its interval includes zero, as expected for designs that count the hours before the reply or count newcomers answered later as controls (Section~\ref{sec:robustness}).

\textbf{Long-term cohort.} The later-period estimates use only newcomers who asked at least a year before the data cutoff. Their 14-day effect (+2.5pp) is larger than that of the full sample (+1.8pp), so the later-period estimates may not carry over to newcomers who joined more recently.

\textbf{Edit quality.} Whether an edit is reverted shows whether other editors accepted it, which is not the same as the quality of the edit; for example, how many editors watch a page also affects whether an edit is reverted. The three groups show how the share of newcomers in each group changes, not which newcomers moved from one group to another.

\textbf{Annotation noise.} Subgroup labels are derived from LLM classification (Cohen's $\kappa$ between 0.68 and 0.90 against each human coder on the 121 test questions). Agreement is lowest for normative questions (Q4, $\kappa$ = 0.68 to 0.75 against 0.885 between the coders on the same questions), so the normative estimate is the most affected. Random misclassification mixes questions of other types into each subgroup and pulls the subgroup estimates toward each other.

\textbf{Reply feature analysis.} Both the regression and the close reading are associational (Section~\ref{sec:reply_features}). Mentors who reply quickly may also differ in other ways, such as mentoring skill, so the associations may reflect the mentor rather than the reply text. The close reading is based on 54 cases and a single coder.

\textbf{Generalizability.} Our data come from English Wikipedia's Growth mentorship system. Results may not generalize to other language editions, other mentorship program designs, or other online communities with different newcomer dynamics.

\subsection{Future Work}

The most immediate next step is to operationalize the ``direct answer'' feature identified in the close reading (whether a reply contains a self-contained answer versus only linking to external resources) through LLM annotation at scale, enabling confirmatory regression analysis across the full treated sample. A causal mediation framework could then decompose the treatment effect into portions attributable to reply speed, content type, and personalization. Randomized experiments, for example A/B testing notification nudges that reduce reply lag, or providing mentors with suggested answer templates for common technical questions, would provide the strongest evidence on which reply features are causally effective. The Growth team's phased rollout also creates an opportunity to estimate the causal effect of mentorship UI access itself, independent of whether a reply occurs. Finally, replicating this analysis across non-English Wikipedia editions and other mentorship platforms (e.g., Stack Overflow, open-source projects) would test generalizability.

\subsection{Ethics and Data Availability}

All data were collected from publicly accessible Wikipedia talk pages and editing logs available through the MediaWiki API. No private or non-public user data were accessed. Usernames appear in the raw data but are not reported in this paper; all examples are paraphrased to prevent identification. Text that the community later removed with \texttt{\{\{redacted\}\}} was not restored.

The full pipeline, from data collection to every table in this paper, is available at \url{https://github.com/everydaylucky/wikipedia-mentorship-icwsm-analysis}, and the input data at \url{https://huggingface.co/datasets/mars1989/wikipedia-mentorship-icwsm-data}.

\section*{Acknowledgments}

We thank Joshua Ashkinaze for his feedback and suggestions throughout this project. We thank Isaac Johnson for discussing the project with us, for comments on the manuscript and for reviewing our code to help make sure the data collection was complete. We thank Martin Urbanec for his detailed answers to our many questions, for helpful discussions, and for his help with data access.

\textbf{Funding.} This work received no external funding. The first author paid all costs of the study, including the fees for the API services, personally.

\textbf{Competing interests.} The authors declare no competing interests.

\textbf{Use of generative AI.} We used DeepSeek-v4-Flash to label question types and Qwen text-embedding-v4 to compute text embeddings, as described in the paper. We used Claude Code to help write the analysis code; the authors reviewed and checked all code, including with automated tests and notebooks that compare processed values with the raw data. We used Claude to polish the wording of the manuscript. The authors take full responsibility for the content of the paper.

\bibliography{references}

\clearpage
\section*{Paper Checklist}

\begin{enumerate}

\item For most authors...
\begin{enumerate}
    \item  Would answering this research question advance science without violating social contracts, such as violating privacy norms, perpetuating unfair profiling, exacerbating the socio-economic divide, or implying disrespect to societies or cultures?
    \answerYes{Yes. Our study analyzes publicly accessible Wikipedia talk page conversations to understand how mentorship affects whether newcomers continue editing. The research aims to improve newcomer onboarding in peer production communities, with no foreseeable negative social impact.}
  \item Do your main claims in the abstract and introduction accurately reflect the paper's contributions and scope?
    \answerYes{Yes. The abstract and introduction state three claims (causal effect of a reply, heterogeneity by question type, reply feature associations) that are each supported by corresponding results sections with appropriate caveats (e.g., the reply feature analysis is explicitly labeled as exploratory and associational).}
   \item Do you clarify how the proposed methodological approach is appropriate for the claims made?
    \answerYes{Yes. The Causal Inference Setup section details the identification strategy, including how newcomer-side differences are controlled with 164 covariates, why random mentor assignment rules out mentor-side confounding, and diagnostics (PS overlap, covariate balance). The reply feature analysis is clearly distinguished as associational.}
   \item Do you clarify what are possible artifacts in the data used, given population-specific distributions?
    \answerYes{Yes. We note that our sample is limited to English Wikipedia newcomers who actively asked a question through the Growth mentorship system, which is a self-selected subset of all newcomers. The Limitations section discusses generalizability constraints.}
  \item Did you describe the limitations of your work?
    \answerYes{Yes. The Limitations subsection discusses the E-value, annotation noise, the associational nature of reply feature analyses, close reading sample size, and generalizability to other platforms and languages.}
  \item Did you discuss any potential negative societal impacts of your work?
    \answerNo{No, because we do not foresee negative societal impacts. The findings concern improving volunteer mentorship for Wikipedia newcomers.}
      \item Did you discuss any potential misuse of your work?
    \answerNo{No, because the findings (e.g., reply faster, answer directly) describe how to improve mentorship quality. We do not see a plausible misuse pathway.}
    \item Did you describe steps taken to prevent or mitigate potential negative outcomes of the research, such as data and model documentation, data anonymization, responsible release, access control, and the reproducibility of findings?
    \answerYes{Yes. The Ethics and Data Availability subsection notes that all data are from public sources, no usernames are reported, and examples are paraphrased to prevent identification. Code is released at \url{https://github.com/everydaylucky/wikipedia-mentorship-icwsm-analysis} and the input data at \url{https://huggingface.co/datasets/mars1989/wikipedia-mentorship-icwsm-data}.}
  \item Have you read the ethics review guidelines and ensured that your paper conforms to them?
    \answerYes{Yes.}
\end{enumerate}

\item Additionally, if your study involves hypotheses testing...
\begin{enumerate}
  \item Did you clearly state the assumptions underlying all theoretical results?
    \answerYes{Yes. The conditional ignorability assumption is stated explicitly in the Identification Strategy section, together with the threats from the newcomer side, the mentor side, overlap and unobserved confounders and how each is addressed.}
  \item Have you provided justifications for all theoretical results?
    \answerYes{Yes. PS diagnostics (overlap, balance), robustness checks (strata sensitivity, trimming, outcome windows, feature ablation, alternative samples, the start of the control window), and a sensitivity analysis (E-values) are provided.}
  \item Did you discuss competing hypotheses or theories that might challenge or complement your theoretical results?
    \answerYes{Yes. We discuss the possibility that unmeasured confounders (e.g., newcomer motivation) could explain results, and report E-values. We also check that the assigned mentor is unrelated to the newcomer and shows no effect on editing outside the reply, which is why mentor-side variables are left out of the PS model (Appendix ``Mentor-Side Variables'').}
  \item Have you considered alternative mechanisms or explanations that might account for the same outcomes observed in your study?
    \answerYes{Yes. The Limitations subsection notes that reply feature associations may reflect mentor quality rather than independent textual effects. The close reading discusses the alternative that newcomer goals (promotion vs.\ learning) rather than reply quality may drive outcomes.}
  \item Did you address potential biases or limitations in your theoretical framework?
    \answerYes{Yes. The Limitations section addresses E-value magnitude, annotation noise, and the associational nature of the reply analysis.}
  \item Have you related your theoretical results to the existing literature in social science?
    \answerYes{Yes. The Related Work section connects our findings to Morrison's (1993) information-seeking taxonomy, prior Wikipedia retention studies, and the causal inference literature on platform interventions.}
  \item Did you discuss the implications of your theoretical results for policy, practice, or further research in the social science domain?
    \answerYes{Yes. The Practical Implications subsection provides three actionable recommendations, and the Future Work subsection outlines next steps including LLM annotation and randomized experiments.}
\end{enumerate}

\item Additionally, if you are including theoretical proofs...
\begin{enumerate}
  \item Did you state the full set of assumptions of all theoretical results?
    \answerNA{NA. No theoretical proofs are included.}
	\item Did you include complete proofs of all theoretical results?
    \answerNA{NA.}
\end{enumerate}

\item Additionally, if you ran machine learning experiments...
\begin{enumerate}
  \item Did you include the code, data, and instructions needed to reproduce the main experimental results (either in the supplemental material or as a URL)?
    \answerYes{Yes. The full pipeline, from data collection to every table, is released at \url{https://github.com/everydaylucky/wikipedia-mentorship-icwsm-analysis}, with the input data at \url{https://huggingface.co/datasets/mars1989/wikipedia-mentorship-icwsm-data}. The remaining raw data (parsed talk pages, revert times, API responses) are rebuilt by the released collection code.}
  \item Did you specify all the training details (e.g., data splits, hyperparameters, how they were chosen)?
    \answerYes{Yes. The LLM prompt was written on 120 human-labeled questions and evaluated on 121 other questions against each human coder. The PS model is an L2-regularized logistic regression with 5-fold cross-fitting.}
  \item Did you report error bars (e.g., with respect to the random seed after running experiments multiple times)?
    \answerYes{Yes. All ATT estimates include 95\% bootstrap confidence intervals (500 resamples).}
	\item Did you include the total amount of compute and the type of resources used (e.g., type of GPUs, internal cluster, or cloud provider)?
    \answerNo{No. The statistical analyses run on CPU nodes of a university computing cluster and need no GPU; LLM annotation and text embeddings used commercial APIs.}
     \item Do you justify how the proposed evaluation is sufficient and appropriate to the claims made?
    \answerYes{Yes. We use multiple outcome measures (6 DVs), multiple robustness checks, and separate within-subgroup PS models. The reply feature analysis uses the full treated sample and all subgroups.}
     \item Do you discuss what is ``the cost'' of misclassification and fault (in)tolerance?
    \answerYes{Yes. The Limitations section discusses how LLM annotation noise (particularly Q4 with lower $\kappa$) could bias subgroup estimates, and the close reading acknowledges its small sample size.}
\end{enumerate}

\item Additionally, if you are using existing assets (e.g., code, data, models) or curating/releasing new assets, \textbf{without compromising anonymity}...
\begin{enumerate}
  \item If your work uses existing assets, did you cite the creators?
    \answerYes{Yes. We cite the Wikimedia Growth team's work (Warncke-Wang et al., 2023), VADER (Hutto \& Gilbert, 2014), ConvoKit (Danescu-Niculescu-Mizil et al., 2013), Perspective API, and all other tools used.}
  \item Did you mention the license of the assets?
    \answerYes{Yes. Wikipedia text is licensed under CC BY-SA 4.0, and our released data use the same license. Our analysis code is released under the MIT License.}
  \item Did you include any new assets in the supplemental material or as a URL?
    \answerYes{Yes. Code is available at \url{https://github.com/everydaylucky/wikipedia-mentorship-icwsm-analysis} and data at \url{https://huggingface.co/datasets/mars1989/wikipedia-mentorship-icwsm-data}.}
  \item Did you discuss whether and how consent was obtained from people whose data you're using/curating?
    \answerYes{Yes. The Ethics and Data Availability subsection notes that all data are from publicly accessible Wikipedia talk pages. Wikipedia contributors agree to public licensing of their contributions upon registration.}
  \item Did you discuss whether the data you are using/curating contains personally identifiable information or offensive content?
    \answerYes{Yes. The Ethics and Data Availability subsection states that usernames are not reported and examples are paraphrased. Perspective API toxicity scores are included as covariates, indicating we are aware of potentially offensive content in the data.}
\item If you are curating or releasing new datasets, did you discuss how you intend to make your datasets FAIR (see \citet{fair})?
\answerYes{Yes. The data are on Hugging Face (\url{https://huggingface.co/datasets/mars1989/wikipedia-mentorship-icwsm-data}) with a dataset card that documents every file, a persistent identifier, an open license (CC BY-SA 4.0) and instructions for reuse with the released code.}
\item If you are curating or releasing new datasets, did you create a Datasheet for the Dataset (see \citet{gebru2021datasheets})?
\answerYes{Yes. A datasheet following \citet{gebru2021datasheets} is released with the dataset (DATASHEET.md).}
\end{enumerate}

\item Additionally, if you used crowdsourcing or conducted research with human subjects, \textbf{without compromising anonymity}...
\begin{enumerate}
  \item Did you include the full text of instructions given to participants and screenshots?
    \answerNA{NA. No crowdsourcing or human subjects research was conducted. The two annotators are co-authors who developed the codebook collaboratively.}
  \item Did you describe any potential participant risks, with mentions of Institutional Review Board (IRB) approvals?
    \answerNA{NA. The study analyzes publicly available data with no direct interaction with participants.}
  \item Did you include the estimated hourly wage paid to participants and the total amount spent on participant compensation?
    \answerNA{NA. No participants were compensated; annotators are co-authors.}
   \item Did you discuss how data is stored, shared, and deidentified?
   \answerYes{Yes. The Ethics and Data Availability subsection addresses this. No usernames are reported in the paper, examples are paraphrased, and the released data are public  Wikipedia records (\url{https://huggingface.co/datasets/mars1989/wikipedia-mentorship-icwsm-data}).}
\end{enumerate}

\end{enumerate}

\clearpage
\appendix

\setcounter{table}{0}
\renewcommand{\thetable}{A\arabic{table}}
\setcounter{figure}{0}
\renewcommand{\thefigure}{A\arabic{figure}}

\section{Appendix}

\begin{table}[h]
\centering
\small
\begin{tabularx}{\columnwidth}{lYl}
\toprule
Outcome & Definition & Type \\
\midrule
\textbf{1+ edit (14d)} & $\geq$1 article edit within 14d & Bin. \\
Number of article edits (14d) & Total article edits within 14d & Count \\
Article edits on 2+ days (14d) & $\geq$2 distinct days with article edits, 14d & Bin. \\
Namespaces edited (14d) & Distinct non-talk namespaces edited, 14d & Count \\
Active days (30d) & Distinct days with a non-talk edit, 30d & Count \\
Reverted (14d) & Any non-talk edit in 14d reverted & Bin. \\
Later periods & 1+ (5+) article edits, days 15--28, 29--60, 61--180, 181--365 & Bin. \\
Month $k$ & $\geq$1 article edit in days $30(k{-}1)$ to $30k$, $k \leq 12$ (figure only) & Bin. \\
\bottomrule
\end{tabularx}
\caption{Outcome definitions. Bold = primary; others supplementary. Article edits are edits in the main namespace; non-talk edits are edits in any namespace except the talk namespaces. Days are calendar days in UTC. Reverted (14d) is shown only in the appendix tables and figures, because it rises with the amount of editing and so does not measure edit quality. }
\label{tab:outcomes}
\end{table}

\begin{table}[h]
\centering
\small
\begin{tabularx}{\columnwidth}{lrY}
\toprule
Block & \# & Examples \\
\midrule
Mentee editing history & 32 & Total edits, revert rate, age \\
Question text features & 37 & Word count, sentiment, politeness \\
Perspective API scores & 8 & Toxicity, insult, profanity \\
Morrison question type & 5 & Q1--Q5 binary dimensions \\
Semantic embeddings & 20 & PCA of text embeddings \\
Temporal fixed effects & 62 & 59 year-month + article context + 2 missing indicators \\
\bottomrule
\end{tabularx}
\caption{Covariate blocks. Full variable definitions in Appendix Table~\ref{tab:covariates_full}.}
\label{tab:covariates}
\end{table}

\begin{figure}[h]
\centering
\includegraphics[width=0.95\columnwidth]{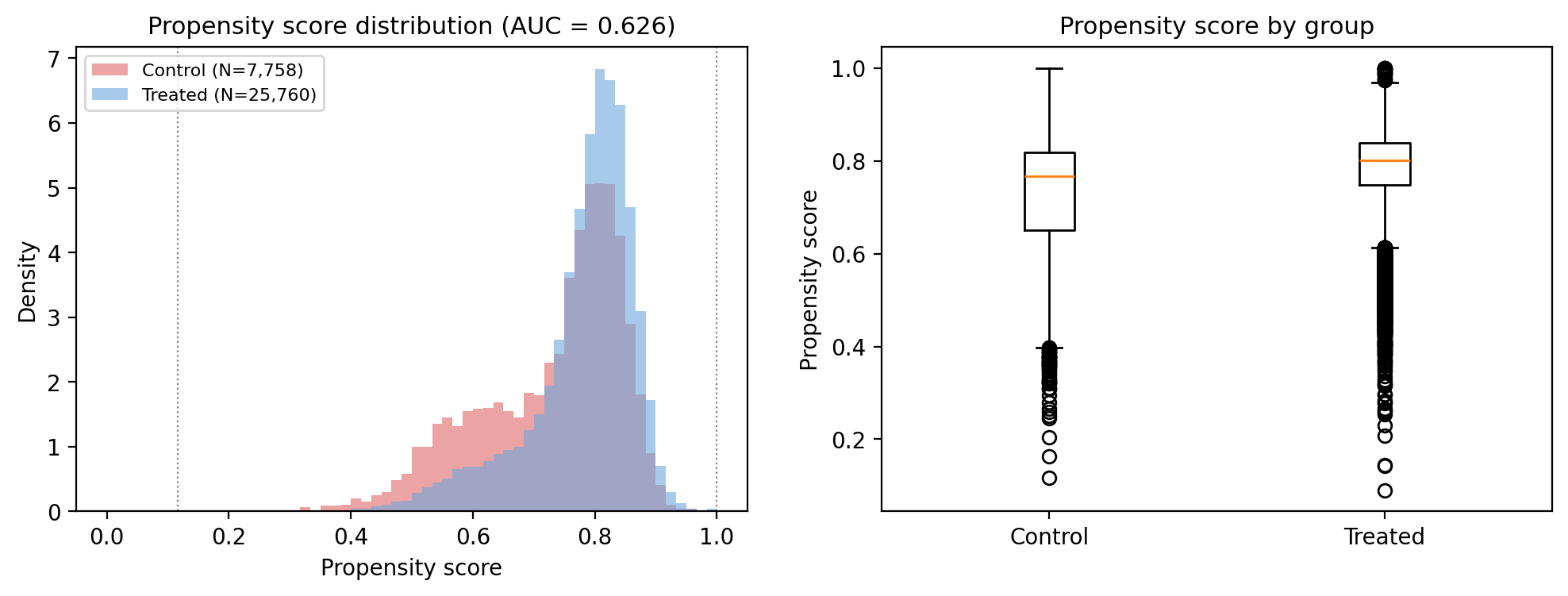}
\caption{Propensity score distributions for treated and control groups with common support region. Only 2 observations (0.01\%) fall outside.}
\label{fig:ps}
\end{figure}

\begin{figure}[h]
\centering
\includegraphics[width=0.95\columnwidth]{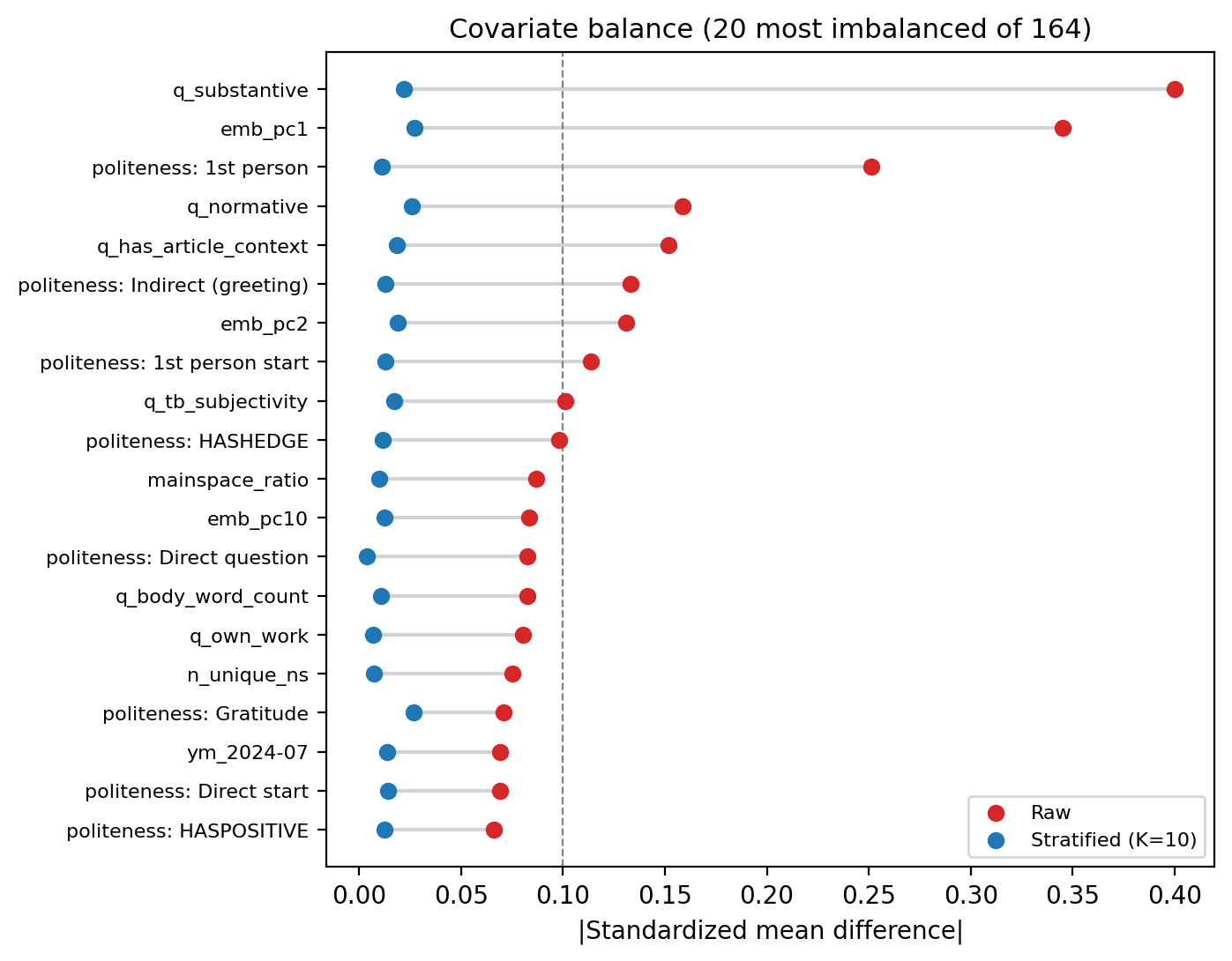}
\caption{Love plot showing the 20 most imbalanced covariates. Red = raw SMD; blue = stratified SMD. Dashed line = 0.1 threshold. After stratification, all 164 covariates fall below the threshold.}
\label{fig:love}
\end{figure}

\begin{table*}[h]
\centering
\small
\setlength{\tabcolsep}{3pt}
\fitcolumn{%
\begin{tabular}{lcccccc}
\toprule
Subgroup & 1+ article edit (14d) & Article edits (14d) & Active days (30d) & Namespaces (14d) & Reverted (14d) & Return within 14d \\
\midrule
Full sample & +.023* [+.011, +.034] & +.92* [+.09, +1.60] & +.16* [+.08, +.25] & +.035* [+.014, +.054] & +.011* [+.002, +.018] & +.021* [+.012, +.029] \\
\midrule
Technical & +.024* [+.007, +.041] & +.61 [$-$.25, +1.46] & +.18* [+.04, +.30] & +.038* [+.008, +.068] & +.008 [$-$.003, +.020] & +.023* [+.010, +.035] \\
Referent (Q2) & $-$.016 [$-$.046, +.012] & $-$.07 [$-$1.31, +1.09] & $-$.09 [$-$.30, +.10] & $-$.009 [$-$.058, +.038] & $-$.020 [$-$.042, +.005] & $-$.004 [$-$.024, +.018] \\
Appraisal (Q3) & +.040 [$-$.012, +.091] & +.50 [$-$5.44, +4.29] & +.26 [$-$.18, +.69] & +.031 [$-$.078, +.140] & $-$.009 [$-$.048, +.027] & +.018 [$-$.026, +.058] \\
Normative (Q4) & +.034* [+.003, +.066] & +2.47* [+.38, +4.79] & +.22 [$-$.08, +.48] & +.049 [$-$.009, +.106] & +.024* [+.002, +.045] & +.027 [$-$.002, +.049] \\
Non-substantive & +.006 [$-$.011, +.021] & +.71* [+.16, +1.36] & +.02 [$-$.08, +.12] & +.006 [$-$.021, +.035] & +.007 [$-$.006, +.020] & +.004 [$-$.007, +.015] \\
Own work (Q5=Y) & +.034* [+.009, +.060] & +1.22 [$-$.34, +2.80] & +.19 [$-$.04, +.41] & +.048* [+.002, +.095] & +.002 [$-$.018, +.018] & +.026* [+.003, +.048] \\
No own work (Q5=N) & +.020* [+.008, +.032] & +.95* [+.27, +1.67] & +.16* [+.07, +.25] & +.033* [+.013, +.054] & +.012* [+.003, +.021] & +.019* [+.010, +.028] \\
\bottomrule
\end{tabular}}
\caption{Within-subgroup ATT for all six outcomes, with 95\% bootstrap intervals; * = interval excludes zero. Same specification as Table~\ref{tab:subgroup} (159 covariates, $K = 5$). Proportions are differences in shares (+.024 = 2.4 percentage points); article edits, active days and namespaces are differences in counts. Q1--Q4 subgroups are not mutually exclusive.}
\label{tab:subgroup_all}
\end{table*}

\begin{table}[h]
\centering
\small
\fitcolumn{%
\begin{tabular}{lrrl}
\toprule
$K$ & ATT & 95\% CI & Sig \\
\midrule
5 & +0.0201 & [+.009, +.032] & * \\
10 & +0.0184 & [+.008, +.030] & * \\
20 & +0.0174 & [+.006, +.029] & * \\
\bottomrule
\end{tabular}}
\caption{Strata count sensitivity.}
\label{tab:strata_sens}
\end{table}

\begin{table}[h]
\centering
\small
\fitcolumn{%
\begin{tabular}{lrlrl}
\toprule
Specification & Feat. & AUC & ATT & Sig \\
\midrule
Full model & 164 & 0.626 & +.018 & * \\
Full $-$ Qtype & 159 & 0.617 & +.020 & * \\
Full $-$ temporal & 102 & 0.617 & +.020 & * \\
Full $-$ embeddings & 144 & 0.619 & +.020 & * \\
Full $-$ Qtext & 127 & 0.624 & +.019 & * \\
Full $-$ Perspective & 156 & 0.626 & +.018 & * \\
Edit history only & 32 & 0.533 & +.031 & * \\
Question feat.\ only & 50 & 0.607 & +.023 & * \\
Qtype only & 5 & 0.595 & +.026 & * \\
\textbf{Full + mentor} & \textbf{170} & \textbf{0.790} & \textbf{+.015} & \\
\bottomrule
\end{tabular}}
\caption{Feature ablation (selected rows of 16 specifications; all but the last are significant). Models with fewer covariates give larger estimates, so the covariates remove part of the selection on newcomer characteristics.  The last row adds the six mentor-side features; the AUC rises because the assigned mentor strongly predicts whether a reply comes, and the ATT is +.015 [$-$.000, +.032]. Appendix~\ref{app:mentor_side} explains why the main model leaves these features out.}
\label{tab:ablation}
\end{table}

\begin{table}[h]
\centering
\small
\fitcolumn{%
\begin{tabular}{lrrl}
\toprule
Trim range & $N$ trimmed & ATT & Sig \\
\midrule
{[}0.01, 0.99{]} & 14 & +0.0187 & * \\
{[}0.05, 0.95{]} & 54 & +0.0190 & * \\
{[}0.10, 0.90{]} & 617 & +0.0205 & * \\
{[}0.15, 0.85{]} & 5,431 & +0.0174 & * \\
\bottomrule
\end{tabular}}
\caption{PS trimming sensitivity.}
\label{tab:trimming}
\end{table}

\begin{table}[h]
\centering
\small
\setlength{\tabcolsep}{4pt}
\fitcolumn{%
\begin{tabular}{lrrr}
\toprule
Outcome & $N$ & Ctrl. & ATT [95\% CI] \\
\midrule
1+ edits, days 15--28 & 32,956 & .079 & \textbf{+.009} [+.000, +.018] \\
5+ edits, days 15--28 & 32,956 & .038 & +.005 [$-$.002, +.010] \\
1+ edits, days 29--60 & 31,665 & .086 & +.005 [$-$.003, +.014] \\
5+ edits, days 29--60 & 31,665 & .040 & +.006 [$-$.001, +.011] \\
1+ edits, days 61--180 & 26,676 & .097 & +.006 [$-$.002, +.017] \\
5+ edits, days 61--180 & 26,676 & .051 & +.006 [$-$.001, +.013] \\
\bottomrule
\end{tabular}}
\caption{Later periods as in Table~\ref{tab:later}, but each row uses all newcomers whose follow-up covers the whole period, so the samples differ between periods. None of the six estimates survives a Benjamini--Hochberg correction. Ctrl.\ = control mean (not reweighted).}
\label{tab:later_all}
\end{table}

\begin{figure}[h]
\centering
\includegraphics[width=0.95\columnwidth]{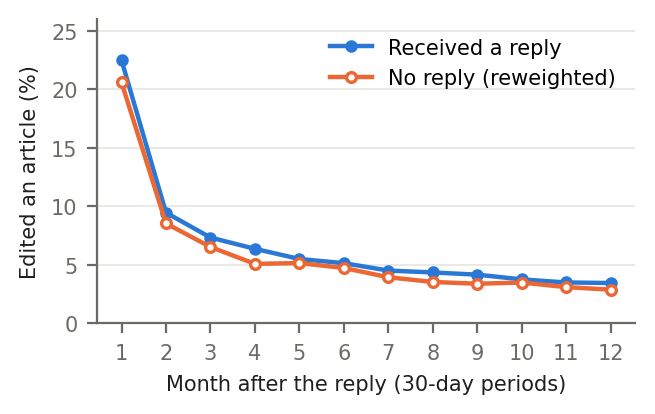}
\caption{Share of newcomers who make at least one article edit in each 30-day month after the anchor, for the cohort of Table~\ref{tab:later}. Controls are reweighted to the treated newcomers' propensity score strata.}
\label{fig:monthly}
\end{figure}

\begin{table}[h]
\centering
\small
\fitcolumn{%
\begin{tabular}{lrrl}
\toprule
Window & $N$ & ATT & 95\% CI \\
\midrule
7 days & 33,793 & +0.0201 & [+.010, +.031] \\
14 days & 33,518 & +0.0184 & [+.008, +.030] \\
21 days & 33,272 & +0.0186 & [+.007, +.030] \\
28 days & 32,956 & +0.0179 & [+.006, +.029] \\
30 days & 32,883 & +0.0166 & [+.004, +.029] \\
60 days & 31,665 & +0.0168 & [+.004, +.029] \\
180 days & 26,676 & +0.0163 & [+.003, +.028] \\
\bottomrule
\end{tabular}}
\caption{Cumulative outcome windows (any article edit within $X$ days), using only newcomers whose follow-up covers the whole window. Longer windows contain the shorter ones, so they cannot show on their own whether the effect lasts; Table~\ref{tab:later} gives non-overlapping periods.}
\label{tab:window}
\end{table}

\begin{figure}[h]
\centering
\includegraphics[width=0.95\columnwidth]{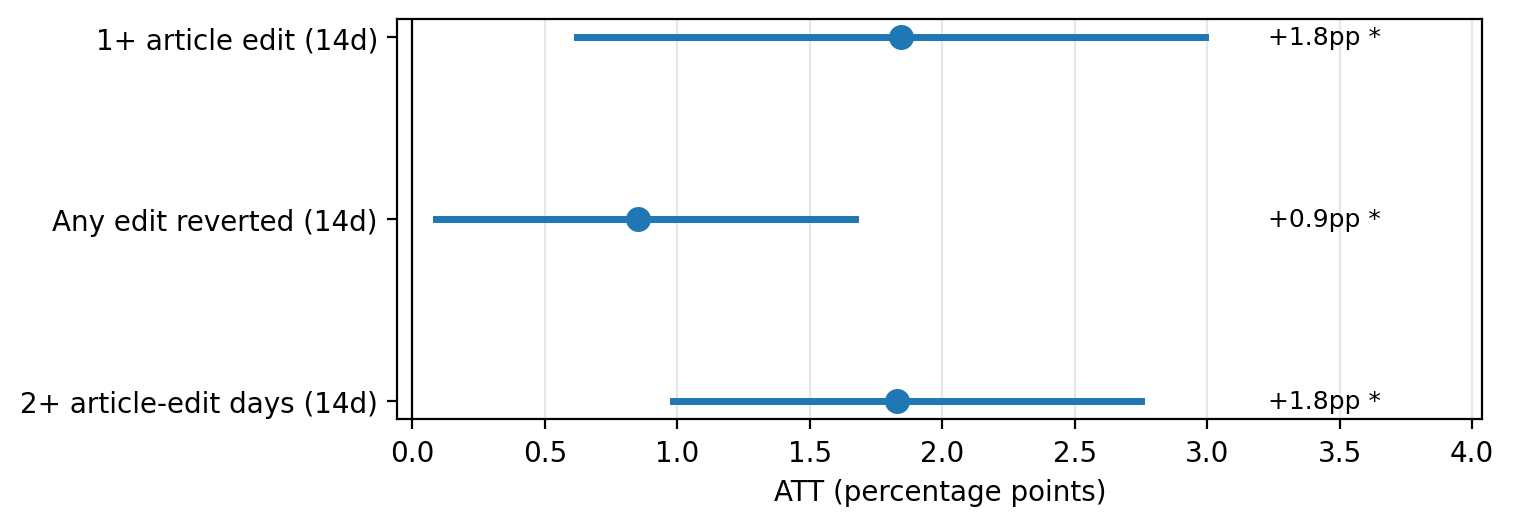}
\caption{Forest plot of ATT estimates across all six outcomes. Blue = significant ($p < 0.05$); gray = non-significant.}
\label{fig:forest_overall}
\end{figure}

\begin{figure}[h]
\centering
\includegraphics[width=0.95\columnwidth]{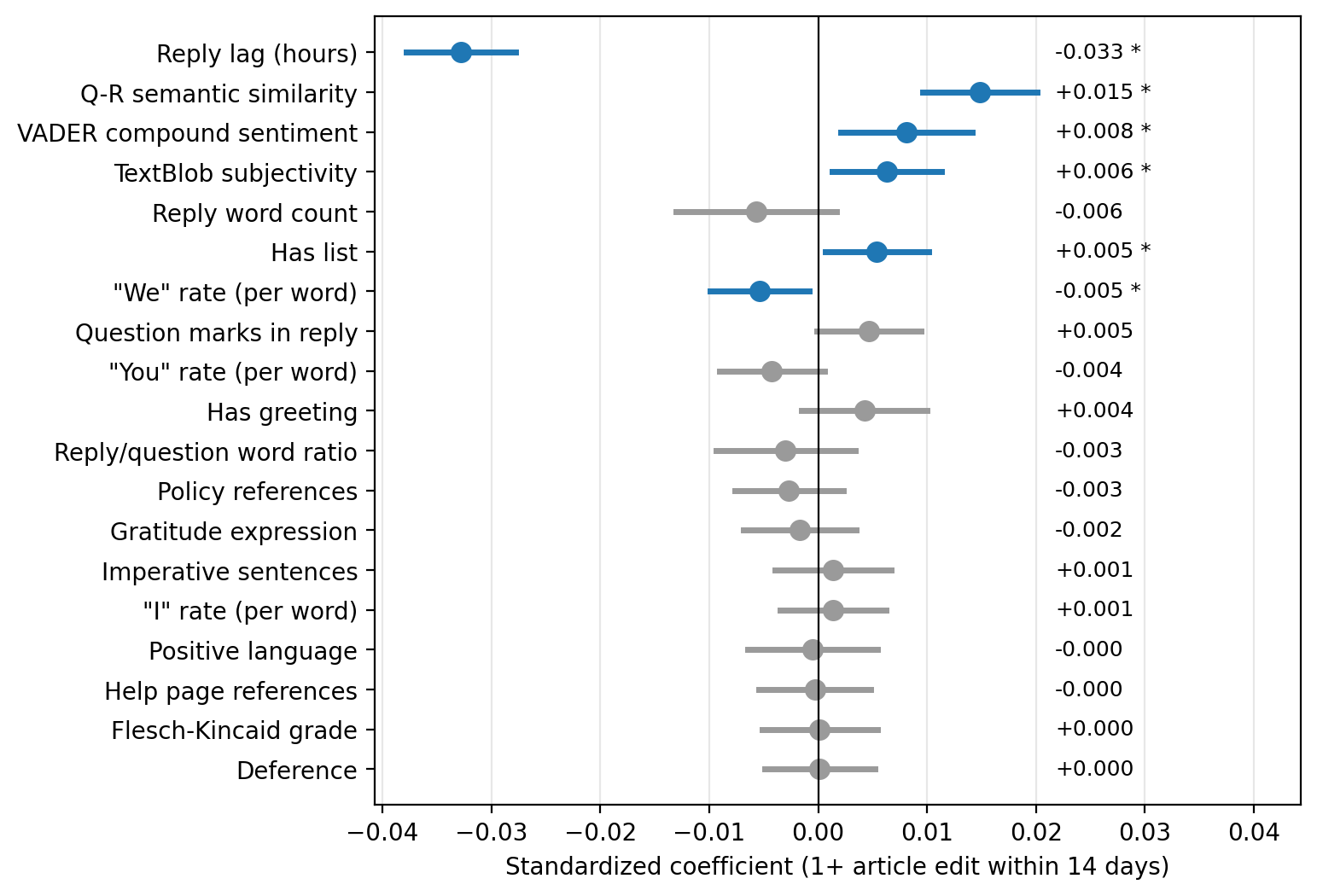}
\caption{Forest plot of the 19 reply feature associations with the primary outcome (95\% CIs). Blue = significant ($p < 0.05$); gray = non-significant. Each $\beta$ represents the change in post-reply editing probability per 1 SD increase in the feature, controlling for 164 pre-treatment covariates.}
\label{fig:reply_features}
\end{figure}

\begin{figure}[h]
\centering
\includegraphics[width=0.95\columnwidth]{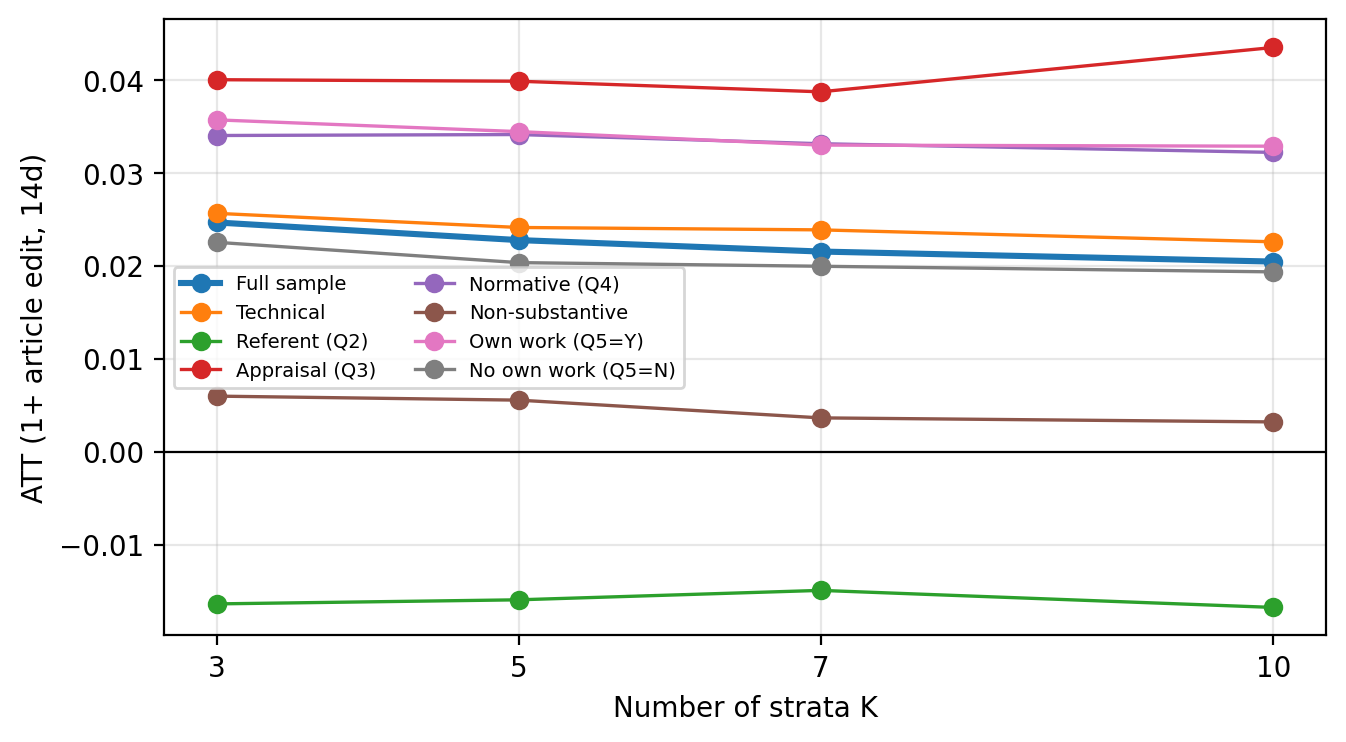}
\caption{ATT stability across strata counts ($K = 3$--$10$) for each subgroup. Significant subgroups keep stable positive ATTs; Referent stays near zero.}
\label{fig:strata_sensitivity}
\end{figure}

\begin{figure}[h]
\centering
\includegraphics[width=0.95\columnwidth]{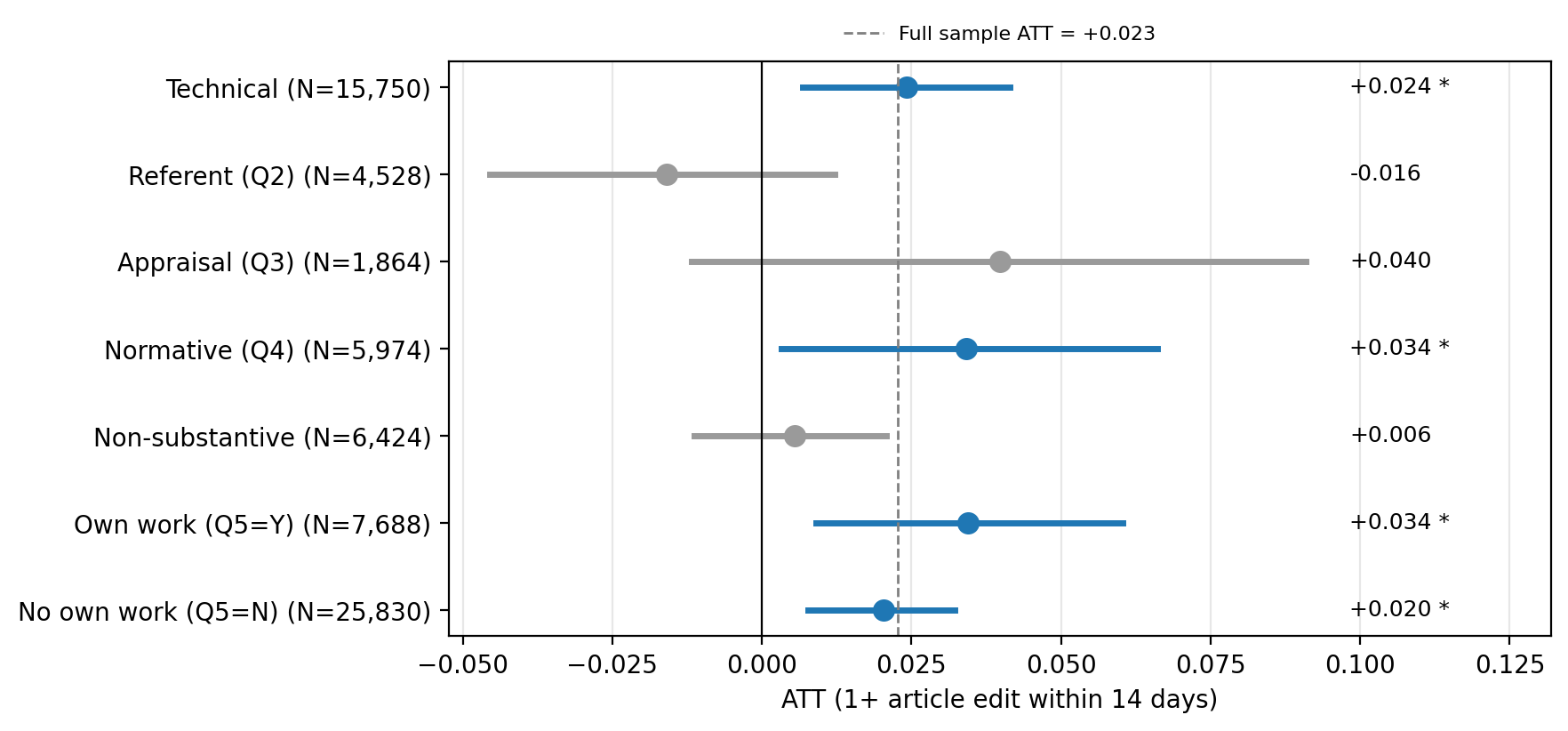}
\caption{Forest plot of within-subgroup ATT estimates for the primary outcome. Blue = significant ($p < 0.05$); gray = non-significant. Dashed line = full-sample ATT with the subgroup specification (+0.023).}
\label{fig:subgroup_forest}
\end{figure}

\begin{figure}[h]
\centering
\includegraphics[width=0.95\columnwidth]{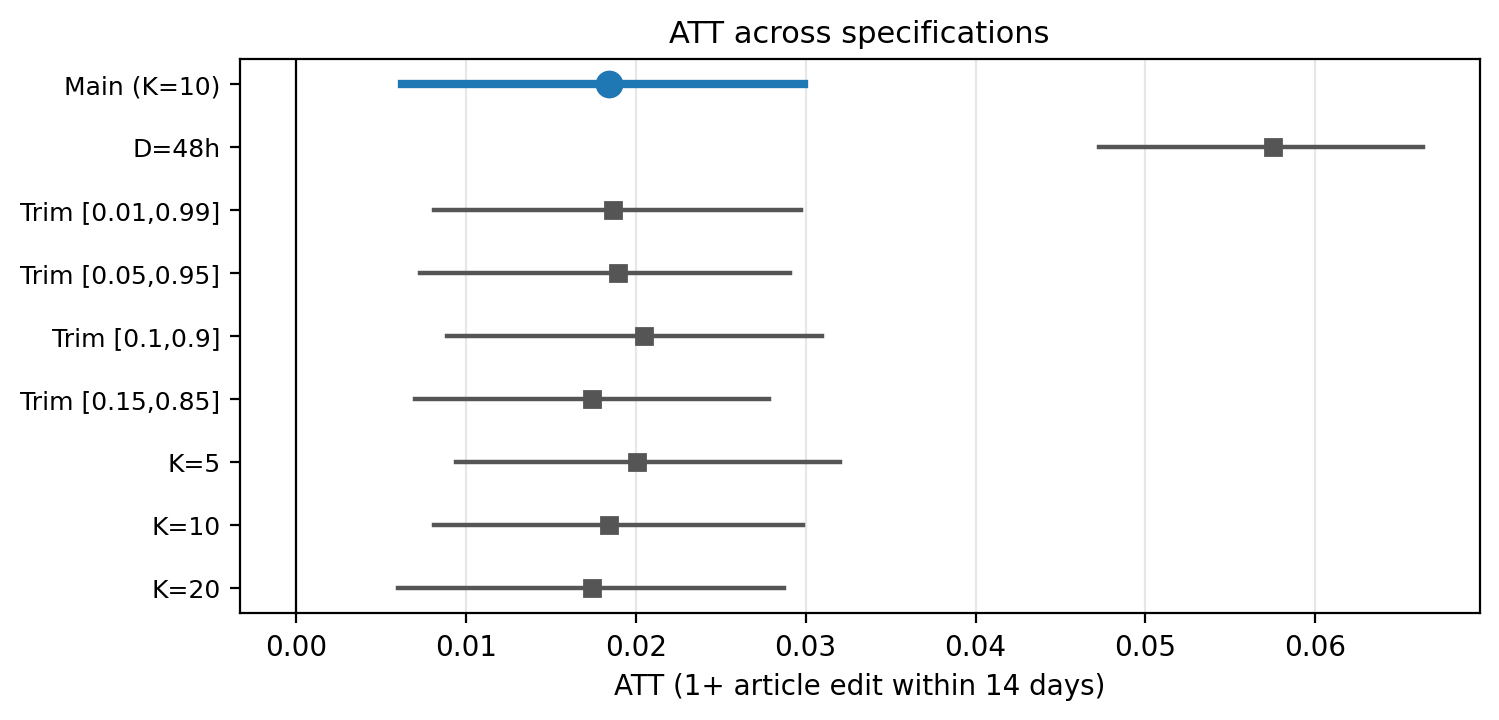}
\caption{ATT across specifications: strata count ($K$), PS trimming, and treatment defined as a reply within 48 hours (see Table~\ref{tab:robust_specs} for why that estimate is larger).}
\label{fig:robustness}
\end{figure}

\begin{table}[h]
\centering
\small
\begin{tabularx}{\columnwidth}{cYr}
\toprule
 & Filter & Newcomers \\
\midrule
& Tagged first questions before the cutoff (one per newcomer) & 37,095 \\
& Account created by another user & $-$452 \\
& Automatically created account & $-$225 \\
& Registered before mentorship began & $-$218 \\
& No account creation record & $-$82 \\
& Claimed by a mentor before asking & $-$19 \\
& Other account exclusions (test, forced local) & $-$5 \\
& Conversation not found & $-$81 \\
& Reply recorded before the question & $-$18 \\
& Question removed in the next edit & $-$841 \\
& Question text hidden or empty & $-$49 \\
& Not in English & $-$1,032 \\
& \textbf{Final sample} & \textbf{34,073} \\
\bottomrule
\end{tabularx}
\caption{Sample construction steps. Among all 42,316 tagged questions, 26.2\% received no reply (22.1\% in 2021, 25.6\% in 2022, 27.9\% in 2023, 26.8\% in 2024, 26.3\% in 2025, 24.3\% in 2026); in the final sample, 23.2\%.}
\label{tab:sample}
\end{table}

\begin{table}[h]
\centering
\small
\setlength{\tabcolsep}{4pt}
\begin{tabularx}{\columnwidth}{Yrrl}
\toprule
Specification & $N$ & ATT & 95\% CI \\
\midrule
Main & 33,518 & +.018 & [+.007, +.030] \\
Assigned mentor replied (vs.\ no reply) & 32,321 & +.019 & [+.008, +.031] \\
No account-type exclusions & 34,419 & +.020 & [+.009, +.031] \\
Excl.\ newcomers whose edits were collected after the cutoff & 31,148 & +.019 & [+.008, +.030] \\
Questions $\geq$30 days before cutoff & 32,949 & +.018 & [+.006, +.029] \\
Incl.\ newcomers without full 14-day follow-up & 34,073 & +.021 & [+.010, +.031] \\
Complete text annotations only & 33,473 & +.020 & [+.009, +.031] \\
\midrule
Reply within 48 hours & 33,518 & +.058 & [+.047, +.066] \\
\bottomrule
\end{tabularx}
\caption{Alternative samples and treatment definitions (primary outcome; each row re-estimates the propensity score). The 48-hour row compares fast replies with late or no replies, a different contrast; its control group includes  newcomers who received a reply later, whose window is anchored at a pseudo-time before their reply arrived, so it is not comparable to the other rows.}
\label{tab:robust_specs}
\end{table}

\begin{table}[h]
\centering
\small
\setlength{\tabcolsep}{3pt}
\resizebox{\columnwidth}{!}{%
\begin{tabular}{llr}
\toprule
\multicolumn{3}{l}{\emph{Who replied (main analysis sample, $N = 33{,}518$)}} \\
(a) & First reply from the assigned mentor & 24,188 \\
(b) & Someone else first, the assigned mentor later & 375 \\
(c) & Only other mentors or editors & 1,197 \\
(d) & No reply & 7,758 \\
\midrule
\multicolumn{3}{l}{\emph{Comparison (treated vs.\ control)}} \\
1 & (a)(b)(c) vs.\ (d): anyone replied (main) & +.018 [+.007, +.030] \\
2 & (a)(b) vs.\ (c)(d) & +.022 [+.012, +.032] \\
3 & (a)(b) vs.\ (d) & +.020 [+.008, +.031] \\
4 & (a)(b) vs.\ (d), from the mentor's first reply & +.019 [+.008, +.030] \\
\bottomrule
\end{tabular}}
\caption{Effect of a reply by who replied (primary outcome; ATT [95\% CI]). Rows 1 to 3 start the treated window at the first reply by anyone; row 4 starts it at the assigned mentor's first reply, which differs only for group (b). Controls start at the pseudo-reply time, except that in row 2 group (c) counts as control and is measured from its actual reply. Row 4 matches ``counting only the assigned mentor's replies'' most closely and is the version reported in Table~\ref{tab:robust_specs}.}
\label{tab:mentor_reply}
\end{table}

\begin{table}[h]
\centering
\small
\setlength{\tabcolsep}{3pt}
\fitcolumn{%
\begin{tabular}{lrll}
\toprule
Outcome & ATT & Fixed strata & Full bootstrap \\
\midrule
1+ edit (14d) & +.018 & [+.006, +.030] & [+.012, +.032] \\
Number of article edits (14d) & +.784 & [$-$.063, +1.53] & [+.174, +1.55] \\
Active days (30d) & +.135 & [+.044, +.229] & [+.068, +.234] \\
Namespaces edited (14d) & +.028 & [+.006, +.049] & [+.013, +.051] \\
Reverted (14d) & +.009 & [+.001, +.017] & [+.001, +.016] \\
Article edits on 2+ days (14d) & +.018 & [+.010, +.028] & [+.012, +.028] \\
\bottomrule
\end{tabular}}
\caption{95\% confidence intervals with the strata held fixed (main analysis) and with a full bootstrap that re-fits the cross-fitted propensity score and re-forms the 10 strata in each of 500 resamples (copies of a newcomer stay in the same fold).}
\label{tab:full_boot}
\end{table}

\begin{table}[h]
\centering
\small
\setlength{\tabcolsep}{3pt}
\fitcolumn{%
\begin{tabular}{lrrrl}
\toprule
Propensity model & AUC & Max SMD & Mean SMD & ATT [95\% CI] \\
\midrule
L2 logit, $C = 1$ (main) & .626 & .027 & .009 & +.018 [+.007, +.030] \\
L2 logit, $C = 0.01$ & .626 & .027 & .008 & +.018 [+.007, +.030] \\
L2 logit, $C = 0.1$ & .626 & .027 & .008 & +.018 [+.007, +.030] \\
L2 logit, $C = 10$ & .626 & .028 & .009 & +.019 [+.007, +.030] \\
Random forest, leaf $\geq 50$ & .620 & .064 & .015 & +.017 [+.006, +.029] \\
Random forest, leaf $\geq 5$ & .616 & .065 & .017 & +.023 [+.012, +.035] \\
Gradient boosting & .617 & .051 & .016 & +.020 [+.009, +.032] \\
\bottomrule
\end{tabular}}
\caption{Propensity score models compared on the main sample ($N = 33{,}518$), each with the same 164 covariates, 5-fold cross-fitting and 10 strata (primary outcome). Max and mean SMD: absolute standardized mean difference over the 164 covariates after stratification. All models use the same 500 bootstrap resamples of newcomers, so the interval of the main model differs slightly from Table~\ref{tab:att}. Changing $C$, the inverse of the regularization strength of the logistic regression, by a factor of 1,000 leaves the AUC at .626, the largest SMD at .027 to .028 and the ATT at +.018 to +.019. Random forests (500 trees, at least 50 or 5 newcomers per leaf) and gradient boosting do not predict better (Brier score .170 for every model, .178 for a constant prediction), leave about twice the imbalance, and give ATTs of +.017 to +.023 with every interval excluding zero.}
\label{tab:ps_models}
\end{table}

\begin{table}[h]
\centering
\small
\begin{tabularx}{\columnwidth}{Yrl}
\toprule
Window start for control newcomers & ATT & 95\% CI \\
\midrule
Random draw from treated lags (main) & +.018 & [+.006, +.030] \\
Mean of 20 random draws & +.017 & [+.007, +.028] \\
Random draw within the PS stratum & +.017 & [+.005, +.028] \\
Mean of 20 draws within the PS stratum & +.017 & [+.007, +.027] \\
Median treated lag (5.7 h) & +.018 & [+.007, +.028] \\
25th-percentile lag (1.0 h) & $-$.011 & [$-$.024, $-$.001] \\
75th-percentile lag (19.2 h) & +.037 & [+.027, +.047] \\
No lag (at the question) & $-$.081 & [$-$.094, $-$.071] \\
\midrule
Both groups from the question & +.009 & [$-$.003, +.021] \\
\bottomrule
\end{tabularx}
\caption{Sensitivity to the start of the outcome window (primary outcome). In the first eight rows treated newcomers are measured from their reply and only the control window changes; the first four draw each control's start from treated reply lags, and the next four give every control the same delay, while treated delays vary. The last row measures everyone from the question ($N = 33{,}561$ with 14 days of follow-up). Appendix~\ref{app:anchor} explains why the random draw is the main specification.}
\label{tab:anchor}
\end{table}

\subsection{Mentor Pool over Time}\label{app:pool}

\textbf{Rollout timeline.} The mentorship module was first deployed on English Wikipedia in March 2021 with help panel routing, and the homepage mentor module followed shortly after. Every new account is assigned a mentor, but the Growth team raised the share of new accounts that could see the homepage module and the help panel entry point in phases: 10\% (March 2022), 25\% (July 2023), 50\% (October 2023), 75\% (February 2025), and 100\% (from February 17, 2025). The team targets about 500 new accounts per active mentor.

\textbf{How we count mentors.} We rebuilt the mentor list from every revision of the mentor list pages before the data cutoff. This gives, for each mentor at each moment, whether the mentor was on the list and whether they were in the automatic assignment pool. Until October 2022 the list was two wiki pages, one for the automatic pool and one for mentors who take manual claims only. After that it is one JSON page, and a mentor is in the automatic pool if the entry says so (\texttt{automaticallyAssigned}) or, where that field is missing, if the mentor's weight is not 0.

\textbf{Number of mentors.} Over the study period, 450 mentors were on the list at some time, and 443 of them were in the automatic pool at some point. At the cutoff, 291 were on the list, 210 in the automatic pool  and 81 taking manual claims only. 175 mentors left the list at least once, and 39 of them came back. The 33,518 newcomers in the main sample were assigned to 400 mentors; the 40 mentors with the most of these newcomers (10\%) account for 34\% of them.

\begin{figure}[h]
\centering
\includegraphics[width=0.95\columnwidth]{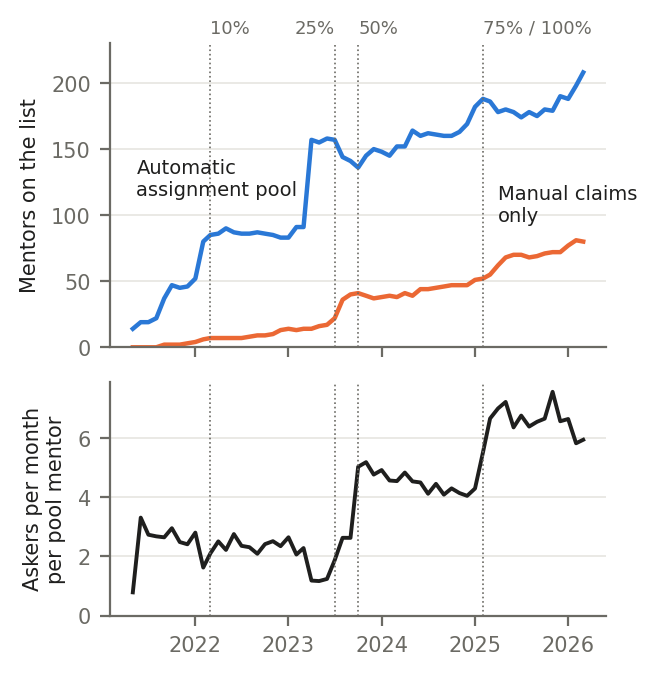}
\caption{Top: mentors on the mentor list at the end of each month, in the automatic assignment pool and taking manual claims only. Bottom: newcomers in the main sample who asked in each month, per mentor in the automatic pool. Dotted lines mark the months in which the share of new accounts that could see the mentorship entry points rose.}
\label{fig:pool}
\end{figure}

\textbf{Pool size.} Figure~\ref{fig:pool} (top) shows the pool at the end of each month. The automatic pool grew from 14 mentors in May 2021 to 85 in March 2022, stayed near that size for a year, rose from 91 to 157 in April 2023, and reached 188 in February 2025 and 208 in March 2026. The Growth team reported 83 mentors in November 2022 and 187 in February 2025 (Phabricator task T323048); our counts for those months are 85 and 188. Mentors who take manual claims only grew from none to 80.

\textbf{Askers per mentor.} Figure~\ref{fig:pool} (bottom) divides the number of sample newcomers who asked in each month by the automatic pool at the end of that month. It counts only newcomers who asked and are in the sample, not every account assigned to a mentor. The ratio rose from 1.2 to 1.9 in July 2023, from 2.6 to 5.0 in October 2023, and from 4.3 to 6.7 between January and March 2025, the months in which the visible share rose. Over the same years, the share of questions without a reply stayed between 22\% and 28\% (Table~\ref{tab:sample}).

\subsection{Mentor-Side Variables}\label{app:mentor_side}

The main PS model uses only newcomer covariates. Leaving out the mentor's reply habits is justified if (1) which mentor a newcomer got is unrelated to the newcomer, and (2) the mentor affects the newcomer's editing only through the reply. We check both with the mentor's past reply rate, the share of earlier questions  on the assigned mentor's talk page that the mentor had answered before this newcomer asked (\texttt{m\_prior\_reply\_rate}, Table~\ref{tab:covariates_full}). We use it when the mentor had at least 5 earlier questions (31,896 of the 33,518 newcomers) and split mentors into thirds by this rate within each month of the question, so that newcomers are compared only with others who asked in the same month. Table~\ref{tab:mentor_side} gives the numbers.

\textbf{(1) Mentor and newcomer.} Newcomers whose mentor was in the top third received a reply 92.7\% of the time, and those whose mentor was in the bottom third 55.5\%. The two groups nevertheless look alike on the 105 covariates of the PS model other than the year-month indicators, with a largest standardized mean difference within month of 0.038 and none above 0.1. 

\textbf{(2) Effects outside the reply.} If the mentor affected editing in other ways, newcomers of responsive mentors would edit more even without a reply, so we look at the 7,521 newcomers who received no reply. In this group the thirds differ, because responsive mentors answer almost every question and the questions they leave unanswered are mostly  those without an editing question or request (largest standardized difference 0.41, for Q1). We therefore regress the primary outcome on the mentor's past reply rate together with the 105 covariates and month indicators. A reply rate 10 points higher goes with $-$0.11pp (95\% CI $-$0.47 to +0.22). Between a bottom-third and a top-third mentor (41 points apart on average) the difference is $-$0.46pp ($-$1.92 to +0.88). Its upper limit is below the effect of a reply (+1.8pp), so a mentor effect outside the reply as large as the effect of a reply is ruled out. Without the covariates, comparing within PS stratum and month, the slope is $-$0.05pp per 10 points ($-$0.54 to +0.35). Intervals come from 500 bootstrap draws of mentors.

The mentor therefore changes whether a reply comes but not who the newcomer is, and we find no effect  of the mentor on editing outside the reply. Mentor-side variables are therefore not confounders, and the main model leaves them out. Adding the six mentor-side features (last row of Table~\ref{tab:ablation}) gives +0.015 [$-$0.000, +0.032].

\begin{table}[h]
\centering
\small
\setlength{\tabcolsep}{4pt}
\fitcolumn{%
\begin{tabular}{lrr}
\toprule
 & Bottom third & Top third \\
\midrule
Past reply rate (mean) & 51.4\% & 91.9\% \\
Newcomers & 10,601 & 10,668 \\
Received a reply & 55.5\% & 92.7\% \\
Max $|$SMD$|$, all & \multicolumn{2}{c}{0.038} \\
Without a reply & 4,714 & 780 \\
Max $|$SMD$|$, no reply & \multicolumn{2}{c}{0.41} \\
\midrule
\multicolumn{3}{l}{Negative control (no reply, $N = 7{,}521$)} \\
Per 10 points (adjusted) & \multicolumn{2}{c}{$-$0.11pp [$-$0.47, +0.22]} \\
Bottom to top (adjusted) & \multicolumn{2}{c}{$-$0.46pp [$-$1.92, +0.88]} \\
Per 10 points (strata) & \multicolumn{2}{c}{$-$0.05pp [$-$0.54, +0.35]} \\
\bottomrule
\end{tabular}}
\caption{Mentor-side checks. Mentors with at least 5 earlier questions are split into thirds of their past reply rate within each question month. SMDs are over 105 covariates, after removing month means. Negative control: change in the primary outcome with the mentor's past reply rate among newcomers without a reply, adjusted for the 105 covariates and month (``adjusted'') or compared within PS stratum and month (``strata''); 95\% intervals from 500 bootstrap draws of mentors.}
\label{tab:mentor_side}
\end{table}

\subsection{Edit Quality}\label{app:quality}

\textbf{When and by whom an edit was reverted.} Wikipedia tags an edit that was later reverted, but the tag does not say when or by whom. For every tagged edit, we searched the page history for a later revision that restored an earlier version of the page or used the undo function.  Of the 16,381 tagged article edits in the 14-day windows, 98.6\% were matched to a reverting revision. Most of the rest are on pages that have since been deleted; we count them as rejected, and counting them as kept gives the same results (Table~\ref{tab:quality_self}). The 48-hour rule requires 48 hours of observation after each edit, so this analysis uses the 33,452 newcomers whose 14-day window ends at least 48 hours before the data cutoff.

\textbf{Newcomers reverting their own edits.} For example, a newcomer changes one part of an article (edit A), notices a minute later that the change broke the formatting, and undoes it (edit B). Edit A is tagged as reverted, but it was reverted by the newcomer, not rejected by other editors. In the 14-day windows, 12,666 article edits were reverted within 48 hours, and 1,375 of them were reverted by the newcomer who made them. These reverts are quick, with a median of 1 minute and 83\% within 10 minutes, compared with  a median of 105 minutes for reverts by other editors. Most are newcomers correcting their own small mistakes or making test edits. We leave out both edit A and edit B, because together they return the page to its earlier state and no other editor judged them. A newcomer whose article edits in the 14 days are all of this kind is treated as having no remaining edit and is placed in group 1; there are 21 such newcomers. Table~\ref{tab:quality_self} shows that counting the edits newcomers reverted themselves as rejected, or as kept, gives the same conclusion.

\textbf{Other cuts.} The main text uses 80\% kept as the cut between groups 2 and 3. Moving this cut from at least one kept edit to 90\% kept (Table~\ref{tab:quality}) raises group 2 by 1.5 to 2.0pp at every cut and does not change group 3 significantly.  Only under the strictest cut, which requires every edit to be kept, does group 3 rise (+0.8pp). Under this cut, a newcomer enters group 3 as soon as one edit is rejected, and newcomers who edit more are more likely to have one edit rejected. A reply makes newcomers edit more, so group 3 grows even if the quality of the edits does not change. This is also why we do not use whether any edit was reverted as a measure of edit quality.

\begin{table*}[h]
\centering
\small
\setlength{\tabcolsep}{3pt}
\fitcolumn{%
\begin{tabular}{lrlrl}
\toprule
 & \multicolumn{2}{c}{Group 2} & \multicolumn{2}{c}{Group 3} \\
Cut (share kept) & Ctrl. & ATT [95\% CI] & Ctrl. & ATT [95\% CI] \\
\midrule
At least one edit & .179 & +.020 [+.010, +.032] & .023 & $-$.001 [$-$.005, +.003] \\
50\% & .171 & +.020 [+.010, +.030] & .031 & $-$.001 [$-$.005, +.004] \\
60\% & .164 & +.020 [+.010, +.029] & .038 & $-$.000 [$-$.005, +.005] \\
70\% & .159 & +.018 [+.009, +.028] & .043 & +.001 [$-$.004, +.007] \\
\textbf{80\% (main text)} & .152 & +.019 [+.010, +.029] & .050 & +.000 [$-$.005, +.007] \\
90\% & .144 & +.015 [+.007, +.025] & .058 & +.004 [$-$.002, +.012] \\
All edits & .129 & +.011 [+.002, +.020] & .073 & +.008 [+.001, +.016] \\
\bottomrule
\end{tabular}}
\caption{Share of newcomers in group 2 (edits kept at or above the cut) and group 3 (below the cut) for different cuts. For ``at least one edit'', group 3 is newcomers who edited but had no edit kept. Group 1 (no article edit) is the same in every row: .798 among reweighted controls, ATT $-$.019 [$-$.031, $-$.009]. $N = 33{,}452$. Ctrl.\ = share among control newcomers reweighted to the treated newcomers' propensity score strata.}
\label{tab:quality}
\end{table*}

\begin{table*}[h]
\centering
\small
\setlength{\tabcolsep}{3pt}
\fitcolumn{%
\begin{tabular}{llrll}
\toprule
Edits reverted & Tagged edits & Rejected & Group 2 & Group 3 \\
by the newcomer & not matched & (controls) & ATT [95\% CI] & ATT [95\% CI] \\
\midrule
\textbf{Left out (main text)} & \textbf{Rejected} & 20.8\% & +.019 [+.010, +.029] & +.000 [$-$.005, +.007] \\
Left out & Kept & 20.6\% & +.019 [+.010, +.029] & +.000 [$-$.005, +.007] \\
Rejected & Rejected & 21.5\% & +.019 [+.010, +.029] & $-$.000 [$-$.006, +.007] \\
Kept & Rejected & 20.4\% & +.018 [+.009, +.028] & +.001 [$-$.005, +.007] \\
\bottomrule
\end{tabular}}
\caption{Results under different treatments of edits that newcomers reverted themselves and of tagged edits not matched to a reverting revision (80\% cut). Rejected (controls) = average share of each newcomer's article edits rejected, among control newcomers who edited.}
\label{tab:quality_self}
\end{table*}

\subsection{Choosing the Start of the Control Window}
\label{app:anchor}

For a treated newcomer, the counterfactual of interest is the same newcomer, at the same moment, without the reply. The control window should therefore start when a reply would have arrived. We assign each control a lag drawn at random from the reply lags of treated newcomers. This gives controls the right start times if two conditions hold.

\textbf{Reply timing does not respond to the newcomer's behavior after asking.} If newcomers who keep editing after they ask were more likely to be noticed and answered, treated newcomers would be selected on post-question activity, and a borrowed start time would be biased. Among the 13,561 newcomers in the sample with no reply in the first 24 hours, those who made an article edit in those 24 hours ($n = 2{,}971$) received a reply in hours 24 to 168 slightly more often (35.4\% vs.\ 31.7\%). Within propensity strata the difference is +0.008 [$-$0.011, +0.028], small and not significant.

\textbf{Reply lags do not depend on who the newcomer is.} A single pool of lags suits every control only if newcomers with different profiles wait about equally long. Reply lags are nearly the same across the ten propensity strata (Table~\ref{tab:lag_strata}), with medians of 4.9 to 6.4 hours, 25th percentiles of about one hour and 75th percentiles of 17 to 21 hours.  Drawing each control's lag only from treated newcomers in the same stratum gives +0.017 [+0.005, +0.028] with one draw and +0.017 [+0.007, +0.027] averaged over 20 draws, close to the main estimate (Table~\ref{tab:anchor}).

\textbf{Why the alternatives are not used as the main estimate.} A fair start time (1) counts only what happens after the reply, or after the moment a reply would have arrived, and (2) observes both groups at the same stage after their question, because newcomers' activity falls with time since asking. The random draw meets both; each alternative breaks one. Fixed lags break (2), because every control starts at the same delay while treated newcomers start  at delays from minutes to days, so the groups are observed at different stages. This holds for any fixed lag, including the median. Measuring both groups from the question breaks (1), because the treated window then includes the hours before the reply arrived and describes  newcomers who will be answered rather than the period after an answer. A landmark $L$ hours after the question (Table~\ref{tab:landmark}) starts both groups together but changes the treatment. Newcomers answered after $L$ count as controls , and the first hours after an early reply are left out, so it measures the value of a reply within $L$ hours compared with a later reply or none, which is a question about reply speed. We report these versions because each relaxes an assumption of the main design, not as alternative estimates of the same quantity. Their estimates are positive except at short fixed lags, and those from common-start designs are about half the main estimate.

\begin{table}[h]
\centering
\small
\setlength{\tabcolsep}{5pt}
\fitcolumn{%
\begin{tabular}{lrrrr}
\toprule
Stratum & Treated & 25th pct (h) & Median (h) & 75th pct (h) \\
\midrule
1 & 1,884 & 0.9 & 5.3 & 17.5 \\
2 & 2,267 & 0.9 & 4.9 & 17.2 \\
3 & 2,573 & 1.0 & 5.4 & 19.9 \\
4 & 2,648 & 1.0 & 6.1 & 20.3 \\
5 & 2,704 & 0.9 & 5.7 & 19.8 \\
6 & 2,734 & 1.0 & 5.7 & 19.3 \\
7 & 2,753 & 1.0 & 6.4 & 20.9 \\
8 & 2,835 & 1.1 & 5.7 & 19.8 \\
9 & 2,854 & 1.0 & 5.8 & 19.4 \\
10 & 2,905 & 0.9 & 5.2 & 17.5 \\
\bottomrule
\end{tabular}}
\caption{Reply lag of treated newcomers by propensity stratum (stratum 1 has the lowest propensity of receiving a reply; strata formed on all 34,073 newcomers).}
\label{tab:lag_strata}
\end{table}

\begin{table}[h]
\centering
\small
\fitcolumn{%
\begin{tabular}{lrrl}
\toprule
Landmark $L$ & $N$ & Treated & ATT [95\% CI] \\
\midrule
12 h & 33,543 & 17,054 & +.003 [$-$.006, +.011] \\
24 h & 33,529 & 20,160 & +.010 [+.002, +.018] \\
48 h & 33,495 & 22,018 & +.008 [$-$.001, +.017] \\
72 h & 33,466 & 22,924 & +.008 [$-$.001, +.016] \\
7 days & 33,311 & 24,361 & +.010 [+.003, +.019] \\
\bottomrule
\end{tabular}}
\caption{Landmark analysis (primary outcome). Treated: a first reply within $L$ hours of the question; controls: no reply within $L$ hours, including newcomers who were answered later. Both groups are measured over the 14 days after the landmark. With controls restricted to newcomers who never received a reply, the 48-hour estimate is +.014 [+.002, +.024].}
\label{tab:landmark}
\end{table}

\begin{table}[h]
\centering
\small
\setlength{\tabcolsep}{3pt}
\fitcolumn{%
\begin{tabular}{lrrlr}
\toprule
Level reached & Treated & Ctrl. & ATT, pp [95\% CI] & $q$ \\
\midrule
\multicolumn{5}{l}{\emph{All edits after the anchor, days 1--365 (main)}} \\
10 edits & 16.0\% & 13.9\% & +2.07 [+0.73, +3.35] & .006 \\
Active (5 in 30 days) & 20.9\% & 17.9\% & +2.98 [+1.53, +4.35] & $<$.001 \\
Very active (100 in 30 days) & 2.9\% & 2.4\% & +0.55 [$-$0.05, +1.10] & .076 \\
500 edits & 1.4\% & 1.0\% & +0.42 [+0.02, +0.79] & .061 \\
\midrule
\multicolumn{5}{l}{\emph{Edits in days 15--365 only}} \\
10 edits & 10.7\% & 9.8\% & +0.91 [$-$0.20, +2.05] &  \\
Active (5 in 30 days) & 12.8\% & 12.3\% & +0.48 [$-$0.74, +1.68] &  \\
Very active (100 in 30 days) & 2.3\% & 1.7\% & +0.58 [+0.04, +1.05] &  \\
500 edits & 1.2\% & 0.9\% & +0.31 [$-$0.07, +0.65] &  \\
\midrule
\multicolumn{5}{l}{\emph{Article edits only, days 1--365}} \\
10 edits & 11.7\% & 10.1\% & +1.59 [+0.38, +2.64] &  \\
Active (5 in 30 days) & 14.7\% & 12.9\% & +1.79 [+0.53, +2.99] &  \\
Very active (100 in 30 days) & 2.2\% & 1.5\% & +0.65 [+0.15, +1.10] &  \\
500 edits & 1.1\% & 0.8\% & +0.29 [$-$0.06, +0.62] &  \\
\bottomrule
\end{tabular}}
\caption{Editing levels reached in the year after the anchor, for the 19,512 newcomers whose anchor is at least 365 days before the data cutoff (14,858 treated). Only edits after the anchor count. 10 and 500 edits are the edit counts Wikipedia uses for autoconfirmed and extended confirmed accounts; we use only the counts, so they do not show whether an account received these rights. Active and very active follow Wikimedia's definitions (at least 5 or 100 edits in a 30-day period, here any 30-day window). Ctrl.\ = control share reweighted to the treated newcomers' propensity score strata. $q$: Benjamini--Hochberg over the four main estimates (2,000 bootstrap draws). The levels were defined in this revision.}
\label{tab:trajectories}
\end{table}

\begin{table}[h]
\centering
\small
\setlength{\tabcolsep}{3pt}
\fitcolumn{%
\begin{tabular}{lrrl}
\toprule
Main type of editing, days 15--365 & Treated & Ctrl. & ATT, pp [95\% CI] \\
\midrule
No edits after day 14 & 76.2\% & 76.4\% & $-$0.13 [$-$1.68, +1.39] \\
Mainly articles & 15.6\% & 14.9\% & +0.75 [$-$0.50, +1.96] \\
Mainly drafts & 0.3\% & 0.3\% & 0.00 [$-$0.24, +0.21] \\
Mainly user and talk pages & 6.4\% & 7.2\% & $-$0.84 [$-$1.80, +0.11] \\
Mainly other namespaces & 0.8\% & 0.8\% & 0.00 [$-$0.33, +0.30] \\
Mixed & 0.7\% & 0.5\% & +0.23 [$-$0.05, +0.49] \\
\bottomrule
\end{tabular}}
\caption{What newcomers edited after day 14 (same newcomers as Table~\ref{tab:trajectories}). A newcomer's main type is the namespace group with at least half of their edits in days 15--365; ``mixed'' if no group has half. Ctrl.\ = control share reweighted to the treated newcomers' propensity score strata. A joint test that the six shares are the same in both groups gives $p = .25$ (bootstrap Wald test on five shares, since the six sum to one).}
\label{tab:editor_type}
\end{table}

\begin{table}[h]
\centering
\small
\fitcolumn{%
\begin{tabular}{lr}
\toprule
Treated newcomers ($N = 26{,}176$) & \\
\midrule
Wrote back after the first reply & 17.4\% \\
Received more than one reply from a mentor & 11.9\% \\
Median turns in the conversation & 2 \\
Median time the conversation stayed active & 8.1 h \\
Asked at least one later question & 8.2\% \\
\quad to the same mentor's page & 7.8\% \\
\midrule
1+ edit (14d) if wrote back / did not & 46.8\% / 17.0\% \\
\quad difference within PS strata & +20.4pp \\
\bottomrule
\end{tabular}}
\caption{Conversation depth and repeated questions. The last two rows use the 25,760 treated newcomers with a full 14-day follow-up. Writing back happens after the reply, so they are associations, not effects.}
\label{tab:conversation}
\end{table}

\subsection{Close Reading of Matched Cases}\label{app:close_reading}

The text features of Section~\ref{sec:reply_features} capture surface-level properties of replies: word count, sentiment, pronoun rates, link counts. But two replies with similar surface statistics can differ in a way that matters. For instance, one directly explains how to perform the task within the reply itself, while the other simply links to an external help page. To explore whether such content-level differences distinguish effective from ineffective replies, we conduct a close reading of matched comparison groups following a most-similar-cases design \cite{seawright2008}.

\textbf{Method.} We identify groups of newcomers who asked the same type of question, had similar pre-treatment profiles, but experienced different post-reply outcomes. We first cluster questions by semantic similarity using HDBSCAN on question embeddings, then select the three largest clusters: citation formatting ($n = 476$), draft submission ($n = 381$), and sandbox use ($n = 352$). Within each cluster, we match newcomers on 34 editing-history features using Mahalanobis distance \cite{mahalanobis1936} (a multivariate metric that accounts for correlations and scale differences across features), requiring pairwise distances $\leq 3.0$, same account-age bracket ($<$1 day / 1--7 days / $>$7 days), and $\leq$3 difference in prior edit count. Each group contains 2 retained newcomers, who made non-talk edits on at least 3 of the 14 days after the reply, and 4 not-retained newcomers, who made no non-talk edit in those 14 days. This definition is stricter than return within 14 days and is used only in the close reading. The 2:4 ratio reflects the low base rate of retention, which limits the number of retained cases available for matching. Reply lag is capped at 7 days. Groups are formed by exhaustive algorithmic search; no selection step involves reply content. This yields 9 groups (54 cases). We code each reply on 8 binary features (Table~\ref{tab:close_reading}), developed iteratively on the first group and applied without modification to the remaining eight. Two features describe the reply's content. \emph{Answers question asked} captures  whether the reply is on-topic (i.e., addresses the newcomer's question rather than providing generic information), while \emph{direct answer in body} captures whether the reply contains the actual answer in the text itself rather than only linking to an external resource where the newcomer must find the answer independently.

\begin{table}[t]
\centering
\small
\fitcolumn{%
\begin{tabular}{lrrl}
\toprule
Reply feature & Retained & Not ret. & $\Delta$ \\
               & ($n$=18) & ($n$=36) & \\
\midrule
Direct answer in body & 61\% & 31\% & +31\% \\
Action steps described & 44\% & 19\% & +25\% \\
Answers question asked & 89\% & 78\% & +11\% \\
Resource annotation & 11\% & 28\% & $-$17\% \\
Follow-up invitation & 28\% & 19\% & +8\% \\
Greeting or personal address & 56\% & 50\% & +6\% \\
Discouraging language & 6\% & 11\% & $-$6\% \\
Links (mean count) & 0.9 & 1.2 & $-$0.3 \\
\bottomrule
\end{tabular}}
\caption{Reply feature prevalence in matched comparison groups (9 groups, 54 cases). ``Retained'' = non-talk edits on $\geq$3 of the 14 days after the reply; ``Not ret.'' = no non-talk edit in those 14 days.}
\label{tab:close_reading}
\end{table}

\textbf{Findings.} The feature that most distinguishes retained from not-retained cases is not reply length or social cues such as greetings; it is whether the mentor answered the question directly in the reply text, rather than pointing the newcomer to an external help page (61\% vs.\ 31\%; Table~\ref{tab:close_reading}). Whether the reply included concrete action steps (e.g., ``surround the reference with \texttt{<ref>} tags'') showed a similar gap (44\% vs.\ 19\%).

For example, in the citation groups, a 25-word reply that showed the inline syntax for a reference tag corresponded to a retained newcomer, while the longest reply in these groups (88 words, 3 links, and a discussion of the newcomer's specific edit history) corresponded to a not-retained one. In this pair, the shorter reply could be acted on without navigating elsewhere, and the longer one could not. In the draft and sandbox groups the pattern was less consistent, with some retained newcomers receiving minimal replies (as short as 5 words) and some not-retained newcomers receiving thorough ones.  In the draft cluster, not-retained newcomers frequently mentioned  creating pages for companies, clients, or organizations, while retained newcomers more often asked about how the drafting process works.

The regression finding that question--reply similarity predicts post-reply editing (Table~\ref{tab:reply_full}) may be related, since on-topic replies could also tend to be more direct, but question--reply similarity measures topical overlap, while ``direct answer'' concerns whether the newcomer can act without navigating elsewhere. Future work could operationalize ``direct answer'' at scale through LLM annotation, enabling a test of whether it accounts for the question--reply similarity association or represents an independent dimension.

\begin{table*}[t]
\centering
\small
\begin{tabularx}{\textwidth}{@{}>{\raggedright\arraybackslash}p{4.2cm}Y@{}}
\toprule
Feature & Definition \\
\midrule
Reply lag (hours) & Hours from the question to the first reply \\
Question--reply similarity & Cosine similarity of the question and reply embeddings (Qwen text-embedding-v4) \\
Sentiment score & VADER compound score \cite{hutto2014}, from $-1$ (most negative) to $+1$ (most positive) \\
Subjectivity & TextBlob subjectivity \cite{loria2020}, from 0 (objective) to 1 (subjective) \\
Share of ``we'', ``I'', ``you'' words & First-person plural, first-person singular and second-person pronouns per word of the reply \cite{pennebaker2003} \\
Reply word count & Number of words in the reply \\
Reply/Q word ratio & Words in the reply divided by words in the question \\
Flesch-Kincaid grade & Readability as a US school grade \cite{kincaid1975} \\
Imperative sentences & Sentences whose root is a base-form verb without a subject, from the spaCy parse \cite{honnibal2020} \\
Has list & The reply contains a bulleted or numbered list (0/1) \\
Question marks & Number of question marks in the reply \\
Has greeting, Gratitude, Positive language, Deference & ConvoKit politeness markers present in the reply \cite{danescu2013} \\
Policy refs & Links to Wikipedia policy or guideline pages \\
Help page refs & Links to pages in the Help namespace \\
Total links, wikilinks & All links, and links to other Wikipedia pages; nearly collinear with the other link counts (variance inflation factor $\approx$ 52) and left out of the models \\
\bottomrule
\end{tabularx}
\caption{Reply features. The 19 features above the last row enter the models; the reply tables use the same names.}
\label{tab:features}
\end{table*}

\begin{table*}[h]
\centering
\caption{Full reply feature associations by subgroup (primary outcome: 1+ article edit, 14d). Each cell shows standardized $\beta$; * $p<.05$, ** $p<.01$, *** $p<.001$. Total links and wikilinks are dropped for collinearity. The main text discusses the significant features.}
\label{tab:reply_subgroup_full}
\resizebox{\textwidth}{!}{%
\begin{tabular}{lrrrrrr}
\toprule
Feature & Technical & Referent & Normative & Non-subst. & Q5=Y & Q5=N \\
 & ($N$=12,659) & ($N$=3,605) & ($N$=4,939) & ($N$=3,943) & ($N$=6,107) & ($N$=19,649) \\
\midrule
Reply lag (hours) & $-$.035*** & $-$.015* & $-$.051*** & $-$.017** & $-$.041*** & $-$.029*** \\
Question--reply similarity & +.017*** & +.005 & +.020** & +.004 & +.011 & +.015*** \\
Sentiment score & +.002 & +.016 & +.010 & +.011 & +.015* & +.007* \\
Share of ``we'' words & $-$.005 & $-$.007 & $-$.006 & $-$.002 & $-$.007 & $-$.005* \\
Subjectivity & +.002 & $-$.006 & +.020* & +.007 & +.016* & +.004 \\
Has list & +.005 & +.006 & +.003 & +.017 & $-$.001 & +.008** \\
Question marks & $-$.000 & +.012 & +.008 & +.001 & +.013* & +.002 \\
Positive language & +.004 & $-$.010 & $-$.005 & $-$.001 & +.003 & $-$.001 \\
Share of ``I'' words & +.004 & +.008 & +.011 & $-$.007* & +.000 & +.001 \\
Share of ``you'' words & $-$.007 & +.004 & +.003 & $-$.004 & $-$.004 & $-$.003 \\
Reply word count & $-$.006 & +.021 & $-$.003 & $-$.013 & $-$.012 & +.000 \\
Reply/Q word ratio & $-$.000 & $-$.017 & $-$.005 & $-$.004 & $-$.031* & $-$.002 \\
Flesch-Kincaid grade & +.003 & +.003 & $-$.003 & $-$.007 & $-$.002 & +.001 \\
Imperative sentences & $-$.002 & +.002 & +.003 & +.013 & $-$.001 & +.001 \\
Has greeting & +.009* & $-$.005 & +.002 & +.001 & +.007 & +.004 \\
Gratitude & $-$.007 & +.012 & $-$.010 & $-$.006 & $-$.004 & $-$.002 \\
Deference & $-$.000 & +.004 & $-$.001 & +.005 & +.006 & $-$.002 \\
Policy refs & $-$.013** & $-$.003 & +.005 & +.008 & +.001 & $-$.005 \\
Help page refs & +.002 & $-$.010 & +.004 & $-$.009 & +.002 & $-$.001 \\
\bottomrule
\end{tabular}}%
\end{table*}

\onecolumn
{\small
\begin{longtable}{llp{7.5cm}}
\caption{Full list of 164 pre-treatment covariates used in the propensity score model, organized by block. Mentor-side features (6 variables, listed separately) are excluded from the main PS model but included in the ablation analysis (Table~\ref{tab:ablation}).}
\label{tab:covariates_full} \\
\toprule
Block & Variable & Description \\
\midrule
\endfirsthead
\toprule
Block & Variable & Description \\
\midrule
\endhead
\midrule
\multicolumn{3}{r}{\emph{Continued on next page}} \\
\bottomrule
\endfoot
\bottomrule
\endlastfoot
\multicolumn{3}{l}{\textbf{Mentee editing history (32 variables)}} \\
& \texttt{n\_edits} & Total edits before question \\
& \texttt{n\_edits\_7d} & Edits in the 7 days before question \\
& \texttt{n\_edits\_1d} & Edits in the 24 hours before question \\
& \texttt{ns0\_mainspace} & Edits to article namespace (ns:0) \\
& \texttt{ns2\_userpage} & Edits to user page namespace (ns:2) \\
& \texttt{ns3\_usertalk} & Edits to user talk namespace (ns:3) \\
& \texttt{ns4\_wp} & Edits to Wikipedia project namespace (ns:4) \\
& \texttt{ns118\_draft} & Edits to draft namespace (ns:118) \\
& \texttt{n\_unique\_ns} & Distinct namespaces edited \\
& \texttt{mainspace\_ratio} & Fraction of edits in mainspace \\
& \texttt{draft\_ratio} & Fraction of edits in draft namespace \\
& \texttt{avg\_sizediff} & Mean byte change per edit \\
& \texttt{std\_sizediff} & Std.\ dev.\ of byte change per edit \\
& \texttt{max\_sizediff} & Largest single byte change \\
& \texttt{neg\_sizediff\_ratio} & Fraction of edits that reduced page size \\
& \texttt{hours\_since\_last\_edit} & Hours between last edit and question \\
& \texttt{active\_span\_hours} & Time between first and last edit \\
& \texttt{account\_age\_hours} & Hours since account creation \\
& \texttt{n\_reverts} & Edits made before the question that had already been reverted before the question \\
& \texttt{revert\_rate\_pre} & \texttt{n\_reverts} divided by \texttt{n\_edits} \\
& \texttt{tag\_visualeditor} & Used the VisualEditor (binary) \\
& \texttt{tag\_mobile} & Edited from mobile (binary) \\
& \texttt{tag\_newcomer\_task} & Used a Growth suggested edit task (binary) \\
& \texttt{tag\_editcheck\_newref} & Triggered the new-reference edit check (binary) \\
& \texttt{tag\_discussion} & Participated in a discussion (binary) \\
& \texttt{log\_create} & Pages created before the question (count) \\
& \texttt{log\_thanks} & ``Thanks'' log entries given or received before the question (count) \\
& \texttt{n\_abuse} & Abuse filter triggers \\
& \texttt{n\_abuse\_warn} & Abuse filter warnings received \\
& \texttt{q\_weekday} & Day of week the question was posted (0--6) \\
& \texttt{q\_hour\_utc} & Hour (UTC) the question was posted (0--23) \\
& \texttt{q\_is\_weekend} & Question posted on Saturday or Sunday (binary) \\
\midrule
\multicolumn{3}{l}{\textbf{Question text features (37 variables)}} \\
& \texttt{q\_body\_word\_count} & Word count of question text \\
& \texttt{q\_body\_char\_len} & Character length of question text \\
& \texttt{q\_vader\_neg/neu/pos/compound} & VADER sentiment scores (4 variables) \\
& \texttt{q\_tb\_polarity} & TextBlob polarity score \\
& \texttt{q\_tb\_subjectivity} & TextBlob subjectivity score \\
& \texttt{q\_has\_sig} & Contains a signature (binary) \\
& \texttt{q\_has\_unsigned} & Contains an unsigned template (binary) \\
& \texttt{q\_link\_count} & Number of wikilinks in question \\
& \texttt{q\_template\_count} & Number of templates used \\
& \texttt{q\_has\_url} & Contains an external URL (binary) \\
& \texttt{q\_mentions\_wp} & Links to Wikipedia: project pages (e.g., WP:RS) \\
& \texttt{q\_mentions\_help} & Links to Help: pages \\
& \texttt{q\_is\_indented} & Question text starts with a talk-page indent (``:'' or ``*'') \\
& \texttt{q\_poly\_*} & 21 politeness strategies from ConvoKit \cite{danescu2013}: Please, Hedges, Deference, Gratitude, Apologizing, 1st/2nd person pronouns, Greeting, Direct question, Positive/Negative language, Subjunctive, Indicative, etc. \\
\midrule
\multicolumn{3}{l}{\textbf{Perspective API scores (8 variables)}} \\
& \texttt{q\_persp\_toxicity} & Toxicity score \\
& \texttt{q\_persp\_severe\_toxicity} & Severe toxicity score \\
& \texttt{q\_persp\_identity\_attack} & Identity attack score \\
& \texttt{q\_persp\_insult} & Insult score \\
& \texttt{q\_persp\_profanity} & Profanity score \\
& \texttt{q\_persp\_threat} & Threat score \\
& \texttt{q\_persp\_sexually\_explicit} & Sexually explicit score \\
& \texttt{q\_persp\_flirtation} & Flirtation score \\
\midrule
\multicolumn{3}{l}{\textbf{Morrison question type (5 variables)}} \\
& \texttt{q\_substantive} (Q1) & Contains a Wikipedia editing question (binary) \\
& \texttt{q\_referent} (Q2) & Does not know what to work on (binary) \\
& \texttt{q\_appraisal} (Q3) & Seeks feedback on existing work (binary) \\
& \texttt{q\_normative} (Q4) & Asks about rules/policies/norms (binary) \\
& \texttt{q\_own\_work} (Q5) & References own prior editing (binary) \\
\midrule
\multicolumn{3}{l}{\textbf{Semantic embeddings (20 variables)}} \\
& \texttt{emb\_pc1}--\texttt{emb\_pc20} & First 20 principal components of 1024-dim question text embeddings (Qwen text-embedding-v4) \\
\midrule
\multicolumn{3}{l}{\textbf{Temporal fixed effects (60 variables)}} \\
& \texttt{ym\_2021-06}--\texttt{ym\_2026-04} & Year-month indicator variables (59 dummies) \\
& \texttt{q\_has\_article\_context} & Question submitted from the help panel while editing an article (binary) \\
& \texttt{*\_missing} & 2 indicators for questions without a Perspective score or an embedding \\
\midrule
\midrule
\multicolumn{3}{l}{\textbf{Mentor-side features (6 variables; excluded from main PS model)}} \\
& \texttt{m\_prior\_mentee\_count} & Mentor's cumulative mentee count before this newcomer \\
& \texttt{m\_prior\_reply\_rate} & Mentor's historical reply rate \\
& \texttt{m\_prior\_avg\_response\_hours} & Mentor's average response time (hours) \\
& \texttt{m\_recent30d\_mentee\_count} & Mentees assigned to mentor in past 30 days \\
& \texttt{m\_recent30d\_reply\_rate} & Mentor's reply rate in past 30 days \\
& \texttt{m\_days\_active} & Days since mentor joined the program \\
\end{longtable}
}

\end{document}